\documentclass[final,5p,times,twocolumn]{elsarticle}

\usepackage{graphicx}

\usepackage{tabularx}
\usepackage{array}

\usepackage[hyphens]{url}
\usepackage[hidelinks]{hyperref}
\hypersetup{breaklinks=true}
\usepackage[dvipsnames]{xcolor}

\usepackage{supertabular}
\usepackage{longtable}

\usepackage{array}\newcolumntype{L}{>{\raggedleft\arraybackslash}X} 

\usepackage{array}\newcolumntype{R}{>{\raggedright\arraybackslash}X} 

\usepackage{array}\newcolumntype{C}{>{\centering\arraybackslash}X} 

\usepackage{xcolor}

\journal{Journal of Systems and Software}

\begin{document}

\begin{frontmatter}

\title{Systematic Literature Review of Validation Methods for AI Systems}
 
\author[1]{Lalli Myllyaho}
\author[1]{Mikko Raatikainen}
\author[1]{Tomi~M\"annist\"o}
\author[1]{Tommi~Mikkonen}
\author[1]{Jukka~K. Nurminen}

\address[1]{University of Helsinki, Finland}

\begin{abstract}
\textit{Context:}
Artificial intelligence (AI) has made its way into everyday activities, particularly through new techniques such as machine learning (ML). These techniques are implementable with little domain knowledge. This, combined with the difficulty of testing AI systems with traditional methods, has made system trustworthiness a pressing issue.

\textit{Objective:}
This paper studies the methods used to validate practical AI systems reported in the literature. Our goal is to classify and describe the methods that are used in realistic settings to ensure the dependability of AI systems.

\textit{Method:}
A systematic literature review resulted in 90 papers. Systems presented in the papers were analysed based on their domain, task, complexity, and applied validation methods.

\textit{Results:}
The validation methods were synthesized into a taxonomy consisting of trial, simulation, model-centred validation, and expert opinion. Failure monitors, safety channels, redundancy, voting, and input and output restrictions are methods used to continuously validate the systems after deployment.

\textit{Conclusions:}
Our results clarify existing strategies applied to validation. They form a basis for the synthesization, assessment, and refinement of AI system validation in research and guidelines for validating individual systems in practice. While various validation strategies have all been relatively widely applied, only few studies report on continuous validation.

\end{abstract}


\end{frontmatter}

\section{Introduction}

Artificial intelligence (AI) has taken society like a storm. Autonomous systems provide us with recommendations of what to watch, where to travel, and where to eat. If newspapers are to be believed, even self-driving cars are just around the corner. With such promises, we must ask how can these systems be trusted? This question has been acknowledged as a major software engineering challenge in the research literature and industry, for both its importance and difficulty \cite{gao2019ai,kumeno2019sofware}. The inherent continuous adaptability and unpredictability of intelligence not only promise opportunities but also create problems unforeseen in traditional software. When considering software \textit{quality in use}, we must pay attention to effectiveness, efficiency, satisfaction, freedom of risk, and context coverage \cite{iso2011iso}. Thus, a self-driving car is supposed to work \textit{as intended} in a crossroads it has never seen, with other cars it has never seen, and with a pothole it has never seen. The question has become even more important with the emergence of easily implementable AI libraries, such as TensorFlow\footnote{https://www.tensorflow.org/} and PyTorch\footnote{https://pytorch.org/}, that provide efficient tools for use even to those who have no experience with or knowledge of the practical implications of the technologies.

In particular, machine learning (ML)---currently a trendy AI paradigm---has spurred roadmaps \cite{breck2017ml} and research (see e.g. \cite{zhang2020machine}) on how learning systems should be and are developed while ensuring their correct functionality. However, earlier research is focused primarily on ML as such and does not address AI systems as a whole where, for example, an ML model is just one component. Furthermore, the main focus has been on the correctness and robustness of the ML models themselves, even though the importance of the correctness and robustness of the entire system or application is both acknowledged by the aforementioned research and is of paramount concern for software engineering overall.

This paper aims to gather a holistic view on \textit{the validation of practical AI systems and applications}. By practical, we mean a complete system that performs a purposeful function in a realistic context. Henceforth, AI systems and applications are referred to shortly as \textit{systems} \cite{iso2011iso}). Our focus on practical systems is motivated by earlier research pointing out the lack of such comprehensive work in current research \cite{zhang2020machine}. A large share of software engineering research has also been observed to be relevant mainly for the research community and less so for practitioners \cite{sjoberg2007future, Ivarsson2011}. Thus, focusing on \textit{practical AI systems} could help to both improve the state of research in general and to allow more practical research to take the leap into actual practice. For clarity, we are not limited to ML but consider any system that can be characterized as containing AI. Respectively, we are not limited to any application domain or task. On the one hand, we are interested in research concerning practical AI systems that are validated using one or more methods, and, on the other hand, research on validation method proposals that are applied to some practical system. Thus, solution proposals, ideas, and theoretical settings---even though often valuable for the research community---are out of the scope of this paper. We understand validation---in contrast to verification---as a method of ensuring that a system works \textit{as intended and designed}, fulfilling its objectives in the context \cite{geraci1991ieee}. This meaning has special indications with practical AI systems, as they are intended to function with little to no human intervention or even to tune their behaviour autonomously in response to the environment \cite{vassev2014autonomy}. Moreover, measures may be needed to ensure that the system keeps fulfilling its objectives well after deployment. Thus, in addition to the initial validation of the system, we are also interested in these additional measures, henceforth referred to as \textit{continuous validation}. Finally, any non-natural entity that autonomously monitors its environment and changes its behaviour accordingly, or is built with ML techniques, may be considered artificially intelligent. However, as the focus of this paper is on practical systems, we are only interested in multi-component systems containing some AI or simpler smart systems, which have clear implications of their usability as is. To avoid confusion, such simple systems will henceforth be referred to as “model-centred”. This means that, for example, autonomous robots and surveillance systems with autonomous access control are in the scope of this paper, whereas whether or not artificial neural networks can be trained to recognize a dog in a stock photo out of context is not.

This study is conducted as a \textit{Systematic Literature Review (SLR)} because SLRs are an efficient means to acquire knowledge about the current stage of research \cite{kitchenham2007guidelines}. SLR as a research method helps to reach beyond the present knowledge on certain domains and venues of individual researchers.
A review based solely on the present body of knowledge may not cover all aspects and possible applications of the subject at hand, especially in the case of such a broad one as AI.
Thus, conducting systematic searches in multiple databases and selecting papers based on inclusion-exclusion criteria expands the scope of the review from what the researcher is already familiar with to a potentially much wider body of knowledge---if the search string formation is successful. SLR can be paralleled to surveys, as both aim to gather evidence from a set of sources. However, these sources differ in respect to research type (SLR or surveys), being either reported research or empirical settings, respectively. SLR was chosen from the two because the existing research---which would provide a good background for the survey---had not been synthesized. We used four databases (Scopus, Web of Science, ACM Digital Library, and IEEE Xplorer) in this SLR to gather the papers using automatic search. The papers were first selected based on their title, keywords, and abstract. Closer selection, sorting, and analyses were conducted after the initial selection.

This paper is organized as follows. Section 2 describes the background and related work. The research questions and methods are described in Sections 3 and 4, respectively. Section 5 provides an overview of the final set of papers, including an analysis of their rigour and relevance. Section 6 summarizes the results for the first research question. Section 7  summarizes the results for the second research question. Sections 8 and 9 discuss our results and their validity, respectively. Section 10 concludes the paper.

\section{Previous work: AI Systems and Validation} \label{sec:back&motiv}

\subsection{Background}

There are numerous definitions for AI. IEEE-USA defines AI in \cite{IEEEUSAAI} as follows: “Artificial Intelligence (AI) is the theory and development of computer systems that are able to perform tasks that normally require human intelligence such as visual perception, speech recognition, learning, decision-making, and natural language processing.” On the other hand, in their report for Stanford University, Stone et al.  \cite{stone2016artificial} use the following: “Artificial Intelligence is that activity devoted to making machines intelligent, and intelligence is that quality that enables an entity to function appropriately and with foresight in its environment.” Based on these, we understand artificially intelligent systems as systems that adapt to, have extensive knowledge of, or learn from their environment or application domain.

The conceptual basis of validation, as applied in this paper, stems from the IEEE standard \cite{geraci1991ieee} that defines validation as “an activity that ensures that an end product stakeholder’s true needs and expectations are met.” In other words, a validation process is for assessing whether or not the final product works as it is supposed to, or whether or not “\textit{the right product was built}.” As such, validation is a part of the verification \& validation (V\&V) process of a system. However, verification and other parts of the V\&V process besides validation are not in the scope of this paper.

The problems in validation begin to stack up when attention is turned to special characteristics of AI systems compared to traditional ones. According to Vassev \& Hinchey \cite{vassev2014autonomy}, if an AI system should function with very few interruptions by humans in its own environment, its requirements should be changed accordingly to include, for example, what objectives the system should perform autonomously, what knowledge it should have, what it should monitor, what it should be aware of, how robust the system is to errors, how adaptable it should be, how dynamic it should be in its adaptations at runtime, how it should resolve unanticipated disruptions, and which parts of the system can be repurposed. Specifying and meeting such highly complex requirements is difficult, as is validating the AI system based on these requirements and showing that the requirements are actually met. Of course, not every AI system is supposed to adapt its behaviour radically after deployment, but surely e.g. a self-configuring system should not only find the initial best configuration but also the most suitable configuration for the changing environment. Additionally, it is a well-known fact that a once descriptive ML model may become outdated as the world around it changes (known as \emph{concept drift} in the literature) \cite{tsymbal2004problem}. Thus, validating AI systems is not only about \emph{what} to validate, but also \emph{when} to validate.

It is important to note that what we mean by validation is not the same as \emph{ML model validation}: ML model validation often means testing the generalization ability of a model \cite{Wang2013}. We do not rule out that ML model validation could be used by some as a validation method for the entire system, but it is not what our study is strictly about.
Therefore, it is not meaningful to discuss the validation of AI systems without using at least nearly realistic systems and contexts. 

Finally, we also differentiate something that we call continuous validation, which emerges from the above-mentioned special characteristics of AI systems.
Continuous validation observes a deployed system and is applied to ensure that the AI system also works in situations unseen during development or in training data, essentially testing that the AI system operates as intended, which is the definition of validation. 
There is, however, no established conceptualization for continuous validation. For instance, Zhang et al.~\cite{zhang2020machine} address continuous validation as a part of online testing, whereas Breck et al.~\cite{breck2017ml} use the term “monitoring testing”. Continuous validation differs somewhat from the traditional way of seeing validation as something usually applied to the end product~\cite{geraci1991ieee}. Many AI systems continue to change their behaviour well after deployment or---as discussed above---are difficult to validate to a satisfactory degree before deployment, thus challenging the idea of an unchangeable end product to a degree. Considering this, post-deployment methods to ensure desired functionality and requirements are met, such as monitoring, fault-tolerance measures, and other safety or quality assurance, are reported as continuous validation in this work.

\subsection{Related work}

As it turns out, validation has been shown to be problematic for AI-based systems. Kumeno's literature review about software engineering challenges for ML systems \cite{kumeno2019sofware} reports that validation was one of the major difficulties reported in the literature related to ML systems. This is paralleled by Gao et al.~\cite{gao2019ai}, who argue that the difficulty of setting the required level of assurance and criteria are amongst the major challenges in ensuring the quality of software utilizing AI. These challenges may arise from many qualities quintessential to AI systems, one of them being the so-called Oracle Problem: if the system is supposed to function autonomously for a long time, adapt, or change its behaviour accordingly to its environment over time, it may be difficult or even impossible to know what the desired outcome will be beforehand. While consistency of behaviour is usually viewed as a desired quality in traditional software, this may not be the case with AI. Another source of challenges includes the sheer scale and diversity of the intended environments and contexts of use. For example, according to Kalra \& Paddock \cite{kalra2016driving}, to statistically significantly prove in a real environment that a fully autonomous car is less prone to fatal accidents than human drivers are would require driving 275 million test miles (approximately 443 million kilometers), and Koopman \& Wagner \cite{koopman2016challenges} present the same kind of infeasibility regarding time. Menghi et al. \cite{menghi2019generating} also argue that even simulations of cyber-physical systems can suffer from the enormous state space of AI systems if not applied purposefully. Consequently, validation challenges have been well observed in the earlier research, and our aim in this paper is to study the validation methods that resolve or alleviate these challenges. 

Gao et al. \cite{gao2019ai} also argue that there is a deficiency in supporting tools for validating AI systems. Readily available tools are not non-existent. For example, Simulink\footnote{https://www.mathworks.com/products/simulink.html} provides tools for simulating cyber-physical systems, and Menghi et al. \cite{menghi2019generating} have developed SOCRaTEs to automatically generate test oracles for Simulink simulations. The Dronology initiative by Cleland-Huang et al. \cite{cleland2018dronology} is another example, which provides an environment for both simulations and physical trials of small unmanned aerial vehicles (UAVs or “drones”). However, Gao et al. do not argue that there are no tools but rather that suitable tools are difficult to find. This could suggest that either the tools and frameworks are still rare or focused on certain domains, or that they are still very recent, scattered, and poorly available. While it is also important to have tools, we focus on how to validate, i.e. the methods for validation that the tools can then support, and do not cover the topic of assisting tools.

The problematic nature of validating AI systems, along with the difficulty of finding suitable solutions, calls out for research on the validation of artificially intelligent systems. In addition to the mentioned tools, reports \cite{zhang2020machine} and guidelines \cite{breck2017ml} have spawned to find answers to these problems in the realm of ML. However, Zhang et al.'s \cite{zhang2020machine} literature review on testing ML software focuses solely on machine learning and mainly on what they call offline testing. This essentially means ML model validation and less the validation of the complete AI system. If we return to the example of a self-configuring system, the ML model-centred approach is questionable: in \cite{nair2017using}, Nair et al. show that even a less accurate learning model outperforms finer learning models when the validation metrics are chosen more appropriately. Thus, validating just the ML model may be inadequate, and the question of \emph{what} is validated should be emphasized. Zhang et al. acknowledge the importance of validating entire systems---which they call “online testing”---and are supported in this by the guideline of Breck et al. \cite{breck2017ml}. Zhang et al. also say that more knowledge on online testing should be attained by the research community. This leaves room and desire for a literature review on the validation methods of practical systems, which is the focus of our paper.

\section{Research questions}

In this work, we seek answers to the following two research questions:

\begin{itemize}
    \item RQ1: What kinds of practical AI systems and their characteristics are presented and validated in the research literature?
    \item RQ2: What kinds of validation methods are used with practical AI systems in the research literature?
\end{itemize}

The goal of RQ1 is to understand the realm of current research. Knowing what kinds of systems are handled in the research literature helps to understand not only what is being covered but also where research gaps may exist. This helps to see and understand the focus, limitations, and shortcomings of the current research. Moreover, a way can be paved for studies focusing on the less covered areas of AI systems.

RQ2 aims to synthesise the scattered information concerning the validation methods, including continuous validation. Knowing what validation methods others have considered effective, what validation methods exist in the first place, and how they are implemented can make a difference when building a new system utilizing AI: skill and awareness of appropriate measures are keys for safe and reliable systems.

\section{Research method}

\begin{figure}[t]
    \centering
    \includegraphics[width=0.9\columnwidth]{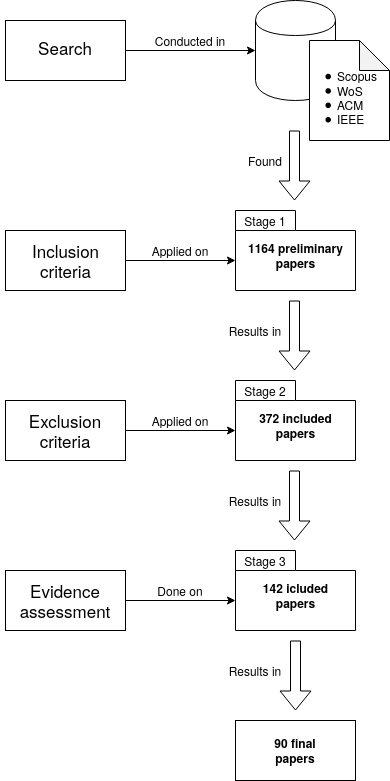}
    \caption{Search and selection process.}
    \label{fig:strategy}
\end{figure}

The research method of this work follows the SLR method  \cite{kitchenham2007guidelines}. The literature search was conducted with four databases (Scopus, Web of Science, ACM Digital Library, and IEEE Xplore) and using semantically the same search string. After the search, three selection stages were applied to reduce the initial set of 1164 papers. The remaining 90 final papers were analysed. The structure of the applied search and selection process is shown in Figure~\ref{fig:strategy}.

\subsection{Search strategy}

The first phase ("Search") of the search and selection process (the full process in Figure~\ref{fig:strategy}) was to gather the initial set of papers by searching scientific databases. The search strategy was revised, and the following steps were taken when conducting the search: \begin{enumerate}
    \item Identify general terms used for systems that utilize ML or other types of AI
    \item Identify synonyms and alternative forms for terms used for systems that utilize AI
    \item Identify prominent domains that utilize AI in the research literature
    \item Identify terms associated with or supporting the term validation as described above
    \item Identify terms that imply usability or industrial relevance of systems described in the research literature
    \item Form a search string based on the identified terms and discussions
    \item If the database allows, use wildcards in the search terms when reasonable
    \item Exclude search hits already found in previous databases
\end{enumerate}

Automatic search was chosen over manual search and snowballing \cite{wohlin2014guidelines} due to the wide range of domains in which AI can be of use. Manual search may lead to missing important venues utilizing the techniques in, e.g., medicine or agriculture. This is especially noteworthy considering the chosen focus on practical systems over the techniques themselves: realistic, applied AI systems may not be presented in venues focusing on, e.g., computer science but in those focusing on the application domain of the said system. For example, a system looking for tumours in CT scans is very likely to be presented in a venue focusing on medicine.

The final search string consisted of three parts: a part for capturing AI; a part to filter out articles that do not emphasize the validation of the system it is presenting; and a part to emphasize articles with the intended use of the system it is presenting. The AI part of the search string consisted of the term “artificial intelligence”, its abbreviation “AI”, a set of terms commonly used for artificially intelligent systems, and a few recent and prominent domains of AI, all connected through logical \texttt{OR}. Inclusion of these domains is not intended to narrow down but rather to widen the scope of the review: according to our earlier experience and acquired knowledge while working on steps 1---3 of the search, these domains are prominent and of interest for this research, but the papers related to them often lack explicit mentions of more general AI terms. We decided it was better to include these few specific terms explicitly, rather than risk not having any of the papers included in the review. These terms may have implications on the results, which are discussed further as a threat to validity in Section~\ref{sec:threat}. The validation part of the search string required the term “validation” and additional terms to emphasize that validation is meant as a part of the V\&V process, as described in Section~\ref{sec:back&motiv}. This was necessary because, as also discussed in Section~\ref{sec:back&motiv}, validation has a different, established meaning in the ML domain, i.e. ML model validation. Thus, combining “machine learning” or “ML” with “validation” without additional restrictions resulted in a very large number of false-positive search hits. The final part was to narrow down the scope to practical systems with intended use or industrial relevance, and it thus consists of terms deemed to reliably imply such qualities in the research literature.

The three parts were connected through the logical \texttt{AND}, and the terms in each part are connected through the logical \texttt{OR}, resulting in the following final search string:

\begin{quote}
\hspace{-1em}\texttt{
\footnotesize{
(AI OR artificial intelligence OR machine learning OR autonomous system OR self-adaptive system OR self-learning system OR intelligent system OR robot OR self-driving vehicle OR autonomous vehicle OR self-driving car OR autonomous car)\\
    \textbf{AND}\\
(validation AND (verification OR testing OR v-model))\\
    \textbf{AND}\\
(user OR customer OR industrial OR industry)
}
}
\end{quote}

The searches were targeted to the title, keywords, and abstract of the articles. However, ACM Digital Library yielded a large number of hits that did not match the search string when searching in these fields. Therefore, in the said database, the search was narrowed down to only include hits in the keywords.

A total of 1164 unique papers were found using the search string. The databases were searched sequentially, starting with Scopus and continuing to Web of Science, then to ACM Digital Library, and finally to IEEE Xplore. This way, 619 papers were found in Scopus, 200 in Web of Science, 57 in ACM, and 551 in IEEE on November 11th 2019, February 10th 2020, March 11th 2020, and March 18th 2020, respectively. The initial set of 1164 papers was reached by removing duplicates.

\subsection{Paper selection}

The paper selection was performed in three stages (Figure~\ref{fig:strategy}) by the first author, who discussed a few unclear papers with the other authors. At the first stage, all 1164 preliminary papers from the search were assessed and included using the inclusion criteria found below. At the second stage, the remaining 372 papers were assessed and excluded using the exclusion criteria. At the third stage, the remaining 142 papers were assessed based on their relevance, resulting in the final set of 90 papers.

\subsubsection{The first stage}

The first stage of selection yielded 372 papers. The selection was based on the title, abstract, and keywords of the papers, and the inclusion criteria below, both of which a paper had to fulfil to be included. 

\begin{itemize}
        \item[IC1]  The paper introduces/discusses a method for validating an AI system \textit{or} the paper applies a method for validating an AI system and 
        \item[IC2]  to count as a system, the presence of at least two components must be declared or implied \textit{or} direct usability of the said system must be declared or clearly implied.
        
\end{itemize}

A consequence of the second criterion (IC2) is that the focus is on papers that acknowledge AI as part of a larger system or as an application as such. For example, the methods of ML model validation or the model's ability to learn are not included in this study \textit{unless} the paper describes other components or if the paper gives a strong indication that the proposed model is being used by someone as is, even if the main focus of the paper is on the AI or ML model. As using a model in a context implies the existence of at least a user interface and raises the model from a technique to a product, this satisfies the definition of a system as a “combination of interacting elements organized to achieve one or more stated purposes” \cite{iso2011iso}.

\subsubsection{The second stage}

The second stage of the selection resulted in 142 papers. The assessment was based on the full texts of the papers. The papers were not included but excluded from the previously narrowed down set. A paper was excluded if at least one of the following exclusion criteria was met: 

\begin{itemize}
    \item [EC1] The full paper was not available in English
    \item [EC2]  The full paper was not acquirable with reasonable effort 
    \item[EC3] The paper turned out to not validate the system it described 
    \item [EC4] The paper turned out to not describe a system or a validation method as described above 
    \item [EC5] The described system turned out to not have properties  recognizable as AI 
    \item [EC6] The paper was less than six pages long 
    \item [EC7] The paper was published in a workshop, a symposium, or  another minor forum
    \end{itemize}

EC1 was enforced by our language skills and the prominent status of English language in SE research. As to EC2, a paper was excluded if it could not be found or accessed in full through measures considered a normal information search. It is hard to assess the effect such papers could have had, as we simply could not get our hands on them, but only a very small number of papers were excluded based on this criterion. EC3---EC5 are derived directly from our research topic. EC6 and EC7 rose from the fact that, according to Kitchenham et al. \cite{kitchenham2010refining}, workshop papers, opinion papers, and other \textit{grey literature} often do not have much to add to the SLRs when included. Kitchenham et al. do advise to be cautious when excluding papers solely based on the publication venue and suggest to first ponder what kind of information is of value to the SLR. However, we are more interested in detailed descriptions of the systems, settings, and validation methods, which, in our experience, are the characteristics that grey literature papers often lack due to the preliminary nature of the results and to page limitations. Also, the preliminary results of workshop and symposium papers are often later expanded to full conference and journal papers. Thus, we decided to focus our effort on papers with presumed higher detailed descriptions, accepting the risk of missing a few relevant grey literature papers.

\subsubsection{The third stage} \label{sec:levelOfEvidence}

The final set of 90 papers was achieved in the third stage of the selection. The selection was made based on the relevance assessment of the papers, so that only papers of high enough relevance were included in the final set. We measure relevance in a paper by the degree of realism in the research settings (see \cite{Ivarsson2011}). The selection was made because our research topic demands a certain level of relevance from the included papers, as we are interested in practical, utilizable systems with intended usability. We decided to label the papers based on the six-label classification of Alves et al. \cite{alves2010requirements} and to outright exclude the lower tier papers. Only papers labelled L4, L5, or L6 were included in this study. The papers were labelled by applying the following criteria for each label:

\begin{itemize}
    \item[L1] No evidence
    \item[L2] The system, its functionality or setup is heavily simplified considering its context, the applicability of the system is not easily recognizable even if it can be recognized as a system, or the evaluation environment is valid but arbitrary considering the context
    \item[L3] The evidence of the paper is not based on empirical evaluation presented in the paper but on a domain expert's preferences of how things should be. For example, general guidelines with no actual evidence presented in the paper
    \item[L4] The system, its functionality or setup is not overly simplified and the evaluation environment of the study is realistic, yet not real considering the context
    \item[L5] The evaluation environment of the study is real but the study is conducted by researchers and the system is not actively used in the industry, even if it is being developed for industrial use
    \item[L6] The described system is actively used by someone in the industry
\end{itemize}

For example, if a robot with obstacle avoidance functionality was tested in just any room with a single box in it, the paper would fall in the L2 category. If the same robot had been tested with a thought-out obstacle course with appropriate structure, it could have fallen in the L4 category. Furthermore, if it had been tested at an industrial facility, it could have been labelled as L5. On the other hand, if it was not only tested in an industrial facility, but also explicitly put into actual, practical use for which it was designed for at that facility, it could be labelled as L6.

\subsection{Information extraction} \label{sec:InformationExtraction}

Data from each paper were extracted using the data extraction form (Table~\ref{dataForm}) and stored on a spreadsheet. Likewise as in the search and selection process, the first author performed the analysis and discussed unclear papers with other authors.

Items F1---F3 in data extraction cover basic bibliographic information. The values for F4---F9 were extracted and generalized from the papers without prior values. That is, the papers were analysed bottom-up by generalizing and categorizing from the data in a manner similar to grounded theory analysis presented in \cite{strauss1998basics}. F4---F7 are characteristics of the AI system under validation in the papers. If a paper did not describe an intelligent system but rather a tool or method to validate such systems, it was classified based on what kind of system was validated with it.

\begin{table}[t]
    \caption{Data extraction form.}
    \centering
 
 \begin{tabularx}{\columnwidth}{>{\centering}p{3cm} X} 
   \hline    Abbreviation & Field \\
    \hline
    F1 & Author(s) \\
    F2 & Title \\
    F3 & Year \\
    F4 & Domain \\
    F5 & Task \\
    F6 & System complexity \\
    F7 & System malfunction impact \\
    F8 & Validation method \\
    F9 & Continuous validation \\
    F10 & Relevance score \\
    F11 & Rigour score \\
    F12 & Machine learning \\
    \hline
    \end{tabularx}
    \label{dataForm}
\end{table}

The domain of a system (F4) is the broad idea of its applicability. For example, the domain of an autonomous car or a system related to it would be “car”, whereas an autonomous radiation shield for projectional radiography would be labelled as “medical”.

The task of the system (F5) is the actual objective it has. For example, the task of a robot searching for and fixing cracks in the hull of a ship would be “maintenance” and the task of a service robot for the elderly would be “care”. Tasks are primarily considered as tasks of the full system. Exceptions to this are papers that focus on describing and evaluating a specific subsystem of a larger system, which has a different task of its own. For example, the objective of an autonomous car is to drive people from one place to another but the objective of the navigation system is to find the most suitable road to the desired destination. Thus, an autonomous navigation system can be seen as a complete subsystem of its own in the context of the car. If a paper described and evaluated the navigation system instead or in the context of a full car, the task would be considered as “navigation” instead of “transportation”, albeit its domain would still be “car”.

The papers were categorized  according to the complexity of the system they presented (F6), which resulted in three groups: \textit{a model-centred system, system,} and \textit{multi-component}. A model-centred system is a simple, yet complete system, which essentially consists of an AI model and a user interface. A system consists of at least two components (not including user-interface) working together. For example, a surveillance system consisting of a camera and an AI model attempting to autonomously recognize unlawful activity would be considered a system per its complexity. A multi-component system is a complex system with multiple components and possibly multiple AI models. For example, an autonomous car would be considered a multi-component system. If a system's complexity was not clear or not inferential within reason, the system was labelled as “unspecified”.

The malfunction impact (F7) of the systems was also considered. Systems were categorized based on their worst yet probable outcome in the event of a malfunction. For example, an autonomous drone for crop ripeness detection could fall out of the sky and kill someone, but that cannot be considered as a probable worst outcome. Instead, such a drone's malfunction impact category would be “economic”, as the probable malfunction would be the misclassification of the crop that would result in the loss of income.

The methods used to validate the system (F8) were extracted. 
Based on the methods used, validation methods were categorized by comparing the similarities and differences of the validation methods each paper presented. These categories were further split into subcategories based on the minor variance between the papers in the same category, resulting in a taxonomy. The names of the categories and subcategories were either directly from or inspired by the papers or the validation methods described. 
In addition, the described methods were analysed qualitatively in greater detail and provided with a descriptive account. 
For each method, at least one high-quality paper is used as a more detailed exemplar, with others supplementing it.
These more detailed descriptions can be found in Section~\ref{sec:analysisVal}.

If a paper describes a method for continuous validation of the system during its life cycle (F9), that too, was analysed. Respectively, as for F8, a categorization, a qualitative analysis, and an exemplar are then presented. These more detailed descriptions can be found in Section~\ref{sec:analysisContVal}.

To assess the quality of the final papers, we adopted the \textit{rigour \& relevance framework} \cite{Ivarsson2011} that represents two orthogonal dimensions for the quality assessment of a paper.
Relevance in the framework refers to the realism of the environment used in the study along with the applied research method. Rather than adopting the relevance measures from the framework directly,  we used the labels of relevance already analyzed during paper selection for the relevance score (F10) (see Section~\ref{sec:levelOfEvidence}) because the labels represent realism similarly as in the framework. We omitted the explicit analysis of research methods for relevance, although the research methods roughly match the applied levels. It is noteworthy that, as relevance was already applied during the selection of the final papers, only the papers assessed to levels 4--6 are included in the set of final papers. 
Rigour (F11) refers to the precision or exactness of the research method used and reported, and it was assessed as a sum according to a rubric of three concerns (context description, design description, and validity discussion of the study) and three scores in each concern adapted from \cite{Ivarsson2011} as follows.

\noindent \textbf{\textit{Rigour scores:}}

\textbf{Context description}
\begin{itemize}
    \item 1: To the degree where a reader can understand and compare it to another context.
    \item .5: Briefly, but not to the degree to which a reader can understand and compare it to another context.
    \item 0: No description.
\end{itemize}

\textbf{Study design description}
\begin{itemize}
    \item 1: To the degree where a reader can understand, e.g., the variables measured, the control used, the treatments, the selection/ sampling used etc.
    \item .5: Briefly, but not in detail.
    \item 0: No description.
\end{itemize}

\textbf{Study validity discussion}
\begin{itemize}
    \item 1: Validity is discussed in detail through several threats or mitigation strategies, or at least through two different validity types in a clear manner.
    \item .5: The validity of the study is mentioned once or twice but not described in detail
    \item 0: No discussion.
\end{itemize}

We assessed whether or not the system utilized ML (F12). If ML or an ML technique was explicitly mentioned, or if the system was interpreted as using ML, the paper was classified as 'yes'. If the paper mentioned other types of intelligence, or performed tasks not typical for ML, the paper was classified as 'no'. Papers not belonging to either of these categories were classified as 'plausible'.

\section{Overview of the final papers}

The final papers are listed in Appendix in Table~\ref{tab:PrimaryStudies}. Results of the data extraction are also provided in the Appendix in Tables \ref{tab:FullData} and \ref{tab:QAData}.

\begin{figure}[h]
    \centering
    \includegraphics[width=0.9\columnwidth]{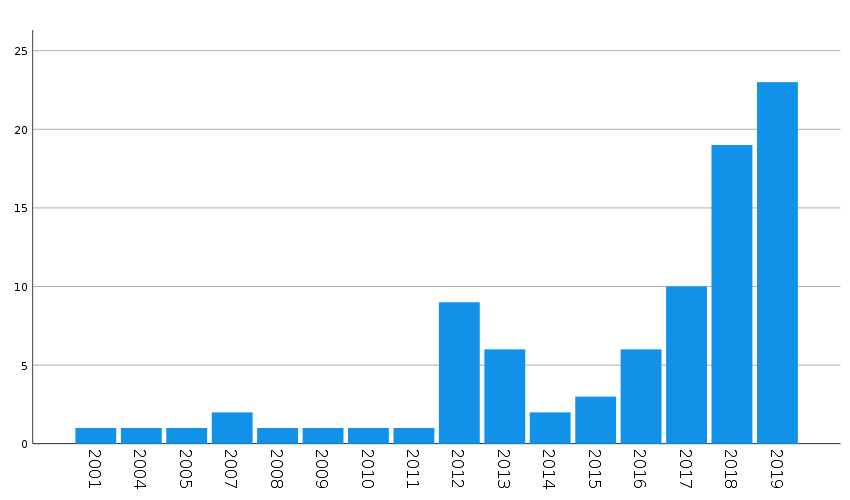}
    \caption{Publication years of papers. 2020 excluded due to incomplete data.}
    \label{publYear}
\end{figure}

Of the 90 final papers, 50 were published by IEEE, 15 by Elsevier, 11 by Springer, 5 by ACM, and 9 by other publishers. The newest papers were published in 2020 and the oldest in 2001, the median being 2017.5. There is a clear rising trend towards recent years (Figure~\ref{publYear}), with 2017, 2018, and 2019 having the most published articles (10, 19, and 23, respectively), as can be expected considering the rising interest in the field. The search covered only the beginning of year 2020. In 2012 and from there onward, a notable rise is seen in the number of papers compared with preceding years.

\begin{figure}[h]
    \centering
    \includegraphics[width=\columnwidth]{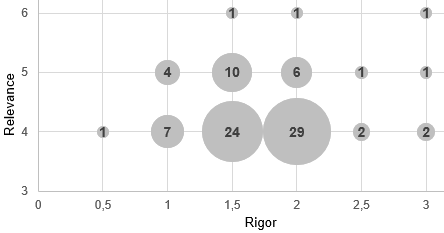}
    \caption{The number of papers at the different levels of rigour and relevance.}
    \label{fig:rigorrelevance}
\end{figure}

On the rigour score (F11), 71 (79\%) papers received a score of 1.5 or 2 out of 3 (see Table~\ref{tab:QAData} in the Appendix, as well as Figure~\ref{fig:rigorrelevance}). This means that most papers either scored high on one or two dimensions (presented in Section~\ref{sec:InformationExtraction}) and poorly on one, or adequately on all. Only seven papers got a rigour score of at least 2.5. Most papers reported well their context and design, whereas result validity was rarely discussed. 62 papers scored a 1 for their context and 55 scored a 1 for their design. In stark contrast, 76 papers do not discuss validity at all.

Relevance was assessed with a relevance level, which was also applied as an inclusion criteria (see Table~\ref{tab:QAData} in the Appendix, as well as Figure~\ref{fig:rigorrelevance}). Thus, only papers at level 4 or higher were included. The majority, i.e., 72.2\% of the papers were assessed as being at level 4, 24.4\% at level 5, and 3.3\% at level 6, which is considered the industrial level.

The highest scoring papers of each validation method are highlighted in their respective subsections in Section~\ref{sec:analysisVal} and Section~\ref{sec:analysisContVal} below. As we already used the relevance score for paper selection, and only papers with a high enough relevance score were included, we consider all papers in the final set to be relevant. Quality comparison and synthesis are therefore primarily based on rigour scores.

\section{Characteristics of validated AI systems (RQ1)} \label{sec:systemsrq}

This section characterizes the validated AI systems by representing quantitative statistics extracted from the papers. The statistics represent results for answering RQ1.

A significant observation is that most presented systems applied some form of ML (F12). Seventy-nine papers either explicitly concerned ML or were reasonably interpretable as such. Six systems were deemed to clearly not be concerned with ML. The final five  were considered plausible. The non-ML papers were not excluded from the results, as the search method was not designed to only include ML papers. However, the reader should be aware of this while going through the results.

Robotics was the most frequent domain, with 21 papers. The second most common domain was systems with some commercial use---e.g. user experience evaluation tools [S88] or storage time estimators [S21]---with 13 papers. The third most common domain was industrial---e.g. a system to assess tunnel construction progress [S38] or a metallic structure defect detection tool [S10]---systems with 12 papers, followed by the car, with 10 papers. A total of 13 domains were recognized in the final papers, excluding unspecified domains. Domains reported with their frequencies are presented in Table~\ref{Dom}.

\begin{table}[h]
    \caption{Domains in the final papers.}
    \centering

\begin{tabularx}{\columnwidth}{>{\raggedright}p{4.3cm} >{\raggedleft}p{2cm} L} 
    \hline    Domain & Number of papers & Percentage \\
    \hline
    Agriculture & 4 & 4.4\% \\
    Aviation & 2 & 2.2\% \\
    Car & 10 & 11.1\% \\
    Commercial & 13 & 14.4\% \\
    Gaming & 1 & 1.1\% \\
    Government & 1 & 1.1\% \\
    Industrial & 12 & 13.3\% \\
    Medical & 7 & 7.8\% \\
    Robotics & 21 & 23.3\% \\
    Safety & 3 & 3.3\% \\
    Smart environment & 4 & 4.4\% \\
    Software testing & 4 & 4.4\% \\
    Unspecified & 3 & 3.3\% \\
    Wearable AI & 5 & 5.6\% \\
        \hline
    \end{tabularx}
    \label{Dom}
\end{table}

Recognition \& classification was by far the most common task reported by the papers, with 31 such systems. It was followed by transportation, with nine papers. Systems with other tasks than these two were presented fewer than seven times each. Eighteen tasks were recognized, excluding unspecified ones. The remaining, less frequent tasks are presented in Table~\ref{tasks}.

\begin{table}[h]
    \caption{Tasks of the presented systems.}
    \centering

\begin{tabularx}{\columnwidth}{>{\raggedright}p{4.3cm} >{\raggedleft}p{2cm} L} 
  \hline
        Task & Number of papers & Percentage \\
    \hline
    Assembly & 1 &1.1\% \\
    Assessment & 6 &6.7\% \\
    Assistance & 1 &1.1\% \\
    Control & 6 &6.7\% \\
    Critical missions & 1 &1.1\% \\
    Decision support & 6 &6.7\% \\
    Design & 1 &1.1\% \\
    Loading & 3 &3.3\% \\
    Maintenance & 5 &5.6\% \\
    Recognition \& classification & 31 &34.4\% \\
    Rehabilitation & 1 &1.1\% \\
    Safety & 4 &4.4\% \\
    Scheduling & 1 &1.1\% \\
    Search and rescue & 1 &1.1\% \\
    Security & 2 &2.2\% \\
    Care & 2 &2.2\% \\ 
    Testing & 3 &3.3\% \\
    Transportation & 9 &10.0\% \\
    Unspecified & 6 &6.7\%\\
        \hline
    \end{tabularx}
    \label{tasks}
\end{table}

The papers represented somewhat evenly the different levels of system complexity. A full system was described in 36 papers. A more complex multi-component system was described in 31 papers. A simple model-centred system was described in 21 papers, and 2 papers were categorized as 'unspecified'.

Considering system malfunction impacts, a total of eight categories were recognized, excluding 'unspecified'. Economic damage was the most common category with 28 papers. Of the systems, 17 were deemed 'lethal', which was the second most common category along with 'nuisance', which also had 17 occurrences. All the malfunction categories can be seen in Table~\ref{tab:critlvl}.

\begin{table}[h]
    \caption{System malfunction impact categories.}
    \centering
 
\begin{tabularx}{\columnwidth}{>{\raggedright}p{4.3cm} >{\raggedleft}p{2cm} L} 
   \hline
    Impact category & Number of papers &Percentage  \\
    \hline
    Bias & 3 &3.3\% \\
    Economic & 28 &31.1\% \\
    Environmental & 1 &1.1\% \\
    Harmful & 8 &8.9\% \\
    Lethal & 17 &18.9\% \\
    Mission critical & 8 &8.9\% \\
    Nuisance & 17 &18.9\% \\
    Privacy & 2 &2.2\% \\
    Unspecified & 6 &6.7\%\\
    \hline
    \end{tabularx}
    \label{tab:critlvl}
\end{table}

Considering how the system domains have been covered over the years (Figure~\ref{fig:domainStream}), commercial, industrial, and medical systems have been presented somewhat consistently. Robots and cars, on the other hand, have significantly risen in numbers in recent years. Other domains pop up here and there, but do not seem to be studied consistently. Perhaps interestingly, the small peak in papers in 2012--2013 seems to consist of nearly equal numbers of systems in almost all domains.

\colorlet{lightyellow}{Yellow!50!}
\colorlet{lightpeach}{YellowOrange!60!}
\colorlet{OwnSpringGreen}{SpringGreen!90!}

\begin{figure}[h]
    \includegraphics[width=\columnwidth]{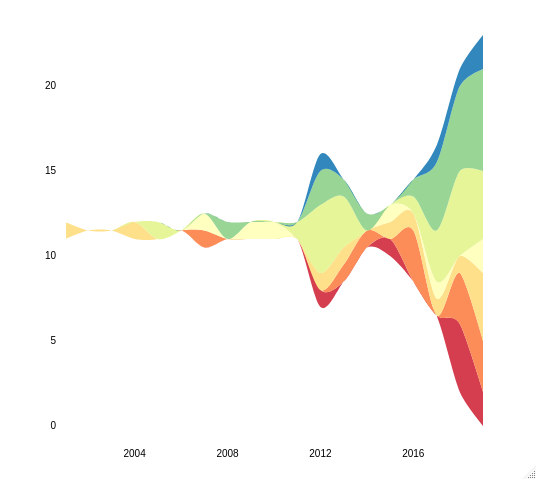}
    \textcolor{NavyBlue}{$\blacksquare$} Wearable AI
    \textcolor{YellowGreen}{$\blacksquare$} Robotics
    \textcolor{OwnSpringGreen}{$\blacksquare$} Other
    \textcolor{lightyellow}{$\blacksquare$} Medical, \newline
    \textcolor{lightpeach}{$\blacksquare$} Industrial, 
\textcolor{Peach}{$\blacksquare$} Commercial, 
    \textcolor{BrickRed}{$\blacksquare$} Car        \caption{System domains over time. Year 2020 has been excluded due to incomplete data.}
    \label{fig:domainStream}
\end{figure}

\section{Validation methods (RQ2)}

\subsection{Quantitative analysis of validation methods } \label{sec:valStat}

This section presents  statistics concerning the validation methods. The categories emerged from the data analysis in a bottom-up manner (Cf. Section~\ref{sec:InformationExtraction}). This section, along with more detailed descriptions of the methods in Sections~\ref{sec:analysisVal}~and~\ref{sec:analysisContVal}, represents results to RQ2.

Of the papers that used only one validation method, a total of 42 used some form of trial (Table~\ref{tab:valMeth}). This is more than twice the number compared to simulation, which was the second most frequent method, with 20 occurrences. In 13 papers, the system was validated with a combination of the listed methods. Expert opinion was rare but was used twice as the sole validation method.

\begin{table}[h]
    \caption{Validation methods used in papers.}
    \centering
\begin{tabularx}{\columnwidth}{>{\raggedright}p{3.5cm} >{\raggedleft}p{2.4cm} L} 
    \hline
    Validation method & Number of papers & Percentage \\
    \hline
    Expert opinion & 2 &2.2\% \\
    Simulation & 20 &22.2\% \\
    Model-centred & 13 &14.4\% \\
    Trial & 42 &46.7\% \\
    Multiple & 13 &14.4\%\\
    \hline
    \end{tabularx}
    \label{tab:valMeth}
\end{table}

If papers using multiple validation methods are partitioned into their respective categories, trial is the largest gainer with an additional 13 occurrences, resulting in a total of 55 occurrences (Table~\ref{tab:valMethErotel}). This means that every combined set of validation methods included a trial of some sort. Simulation gained 11 occurrences, rising to a total of 31 occurrences. Model-centred validation and expert opinion gained 3 and 1 occurrences, resulting in a total of 16 and 3 occurrences, respectively.

\begin{table}[h]
    \caption{Individual validation methods used when multiple methods are divided into their respected categories.}
    \centering
\begin{tabularx}{\columnwidth}{>{\raggedright}p{3.5cm}  L} 
\hline
    Validation method & Occurrences \\
    \hline
    Expert opinion & 3 \\
    Simulation & 31 \\
    Model-centred & 15 \\
    Trial & 55 \\
    \hline
    Total & 104 \\
    \hline
    \end{tabularx}
    \label{tab:valMethErotel}
\end{table}

Methods for continuous validation (described in more detail in Section~\ref{sec:analysisContVal}) over the life cycle of the presented system were described in 14 papers (Table~\ref{tab:contVal}). This leaves 76 papers that did not report using any method of continuous validation. The papers described a total of six methods for continuous validation. Safety channel was the most frequent one with five occurrences. Safety channel was followed by a failure monitor with four occurrences. The other described methods were input and output restrictions, redundancy, and voting.

\begin{table}[h]
    \caption{Methods for continuous validation.}
    \centering
\begin{tabularx}{\columnwidth}{>{\raggedright}p{3.5cm} >{\raggedleft}p{2.4cm} L} 
        \hline
    Validation method & Occurrences & Percentage \\
    \hline
    Failure monitor & 4 &4.4\% \\
    Input restrictions & 2 &2.2\% \\
    None & 63 &70\% \\
    None \--- method description & 13 & 14.4\% \\
    Output restrictions & 2 &2.2\% \\
    Redundancy & 1 &1.1\% \\
    Safety channel & 5 &5.6\% \\
    Voting & 1 &1.1\%\\
    \hline
    \end{tabularx}
    \label{tab:contVal}
\end{table}

When observing the popularity of the validation methods in different domains (Table~\ref{tab:domainsmethods}), trials appeared to be popular in all domains. Model-centred approaches were relatively popular in commercial, industrial, and medical domains---all of which have raised research interest quite consistently over time (cf. Section~\ref{sec:systemsrq}). Simulations, in turn, were extremely popular in cyber-physical domains, especially in cars and robots.

\begin{table}[h]
 \caption{The validation methods used in each domain.}
    \centering
\begin{tabularx}{\columnwidth}{>{\raggedright}p{2cm} L L L L} 
        \hline
    Domain & Expert opinion & Simulation & Model-centred & Trial \\
    \hline
    Agriculture & 0 & 0 & 0 & 4 \\
    Aviation & 0 & 1 & 1 & 1 \\
    Car & 0 & 7 & 1 & 5 \\
    Commercial & 0 & 1 & 4 & 8 \\
    Gaming & 0 & 0 & 0 & 1 \\
    Government & 0 & 1 & 0 & 0 \\
    Industrial & 1 & 2 & 5 & 7 \\
    Medical & 0 & 1 & 3 & 4 \\
    Robot & 0 & 13 & 0 & 14 \\
    Safety & 0 & 1 & 1 & 1 \\
    Smart \newline environment & 0 & 2 & 1 & 1 \\
    Software testing & 0 & 0 & 0 & 4 \\
    Unspecified & 1 & 1 & 0 & 1 \\
    Wearable AI & 1 & 0 & 0 & 4 \\
    \hline
    \end{tabularx}
    \label{tab:domainsmethods}
 \end{table}

\begin{figure}[h]
    \centering
    \includegraphics[width=\columnwidth]{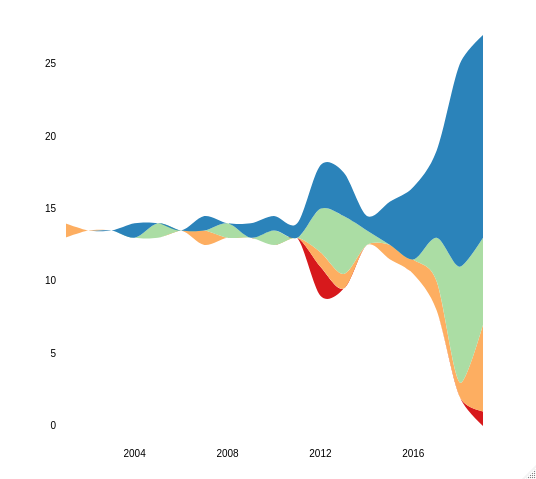}
    \raggedright
    \textcolor{NavyBlue}{$\blacksquare$} Trial
    \textcolor{YellowGreen}{$\blacksquare$} Simulation, 
\textcolor{Peach}{$\blacksquare$} Model-centred, 
    \textcolor{BrickRed}{$\blacksquare$} Expert opinion,    \caption{Validation methods over time. Year 2020 has been excluded due to incomplete data.}
\label{fig:valMethodStream}
    
\end{figure}

Considering how the popularity of validation methods has evolved over time (Figure~\ref{fig:valMethodStream}), trials (blue) and simulations (green) have nearly exploded in recent years, following the overall increase of research (cf. Table~\ref{publYear}). This change in popularity, especially in the case of simulations, follows the rising interest in cyber-physical domains such as cars and robots, in which simulations and trials were very popular (cf. Table~\ref{tab:domainsmethods} and Section~\ref{sec:systemsrq}. Model-centred approaches (orange) and expert opinions (red) have been used quite consistently, although they remain low in total numbers without any clear trends.

\subsection{Descriptive analysis on validation methods} \label{sec:analysisVal}

\begin{figure}[h]
    \centering
    
    \includegraphics[width=\columnwidth]{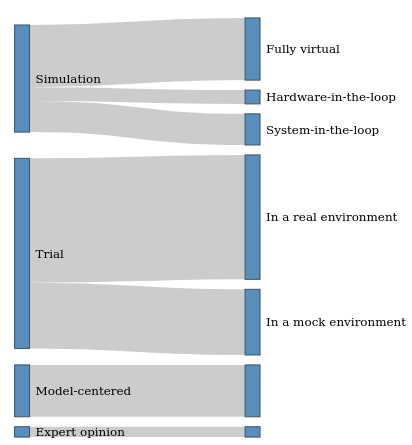}
    \caption{The taxonomy of the validation methods, along with their relative commonness represented by the height of the bar.}
    \label{fig:taxonomy}
\end{figure}

While analysing the papers, we observed that the validation methods form a shallow taxonomy (Figure~\ref{fig:taxonomy}). In this section, we describe in more detail how the validation methods within the taxonomy are used and which factors seem to affect the choice of using these methods. We go over the basic principles of the four main validation methods (cf. Table~\ref{tab:valMeth}), and their variations, along with their strengths and weaknesses. This section, along with the following Section~\ref{sec:analysisContVal}, provides a more in-depth analysis for the methods presented in Section~\ref{sec:valStat} for RQ2. The papers that used multiple validation methods or multiple variations are acknowledged individually in each section.

\subsubsection{Simulation} \label{simulation}

In this section, we describe three simulation variations. Overall, in simulations, the system is validated in a clearly artificial environment (virtual or real) mimicking the actual deployment environment. Simulations seem to be heavily favoured in cyber-physical systems, as 11 out of 21 papers about robotics and 8 out 10 papers about cars describe simulations (see Table~\ref{tab:domainsmethods}).

\textbf{Fully virtual simulation.}
In a fully virtual simulation, the final deployment environment of the system is replicated with a virtual simulator, along with possible physical components of the system itself [S1], [S5], [S9], [S20], [S22], [S34], [S35], [S37], [S38], [S44], [S50], [S53], [S56], [S70], [S73], [S75], [S77], [S78]. Basically, the inputs of the real scenario are replaced with inputs created by the simulator. Depending on the system, the outputs can be assessed as is or they can be mimicked by a virtual model of the system. 11 of the 18 systems are either cars or robots. The only paper with a rigour score of 2.5 or higher is [S78]. In [S78], the test cases are modelled real-life traffic situations. A single vehicle in the situation is replaced with a virtual model of an autonomous car while other models in the simulator mimic the vehicles of the modelled situation. The system-under-test navigates amidst the models of real cars, and the simulator observes whether or not the virtual model violates declared safety constraints and reports on the violations.

Alternatively, the test cases can, for example, be passed down as ontologies or as a set of constraints from which the simulator extracts the different situations in a combinatorial manner [S1], [S9], or the tests can be designed by hand [S9]. Another example is to inject faults and errors into the software or some established test scenarios to find the components and test cases that are the most prone to faults [S5].

Fully virtual simulation makes validating safety-critical systems safer, as the functionality of the system can be tested in potentially dangerous situations without actual danger [S78], [S1]. Also, simulations provide a faster and more convenient way to run the tests of a highly complex system than a trial in a real environment would [S1], [S9], [S78].

As a downside, a fully virtual simulation puts a lot of pressure on the simulator [S1]. Any deviation between the real environment and the simulation or the full system and its simulated counterpart can have serious consequences. That is, as also noted by [S78], the validity of a simulator is difficult to prove.

Simulation can also serve as initial validation before other validation methods [S20], [S22], [S75]. This way the full system does not have to be built before there is some level of certainty of its functioning through modelling the system. As such, some potentially malfunctioning components will fail rapidly and can be fixed early on.

\textbf{Hardware-in-the-loop simulation (HIL).}
In HIL simulation---as it is called by [S6]---the final deployment environment of the system is replicated with a virtual simulator or some other artificial means [S6], [S72], [S83], [S86]. The main difference compared with the fully virtual simulation is that other, non-virtual components of the system are included in the validation setup for handling inputs or outputs. All systems in this category are cyber-physical systems (cars and robots). No paper reached a higher rigour score than 2.0 on the rigour scale, and so [S72] is used as the top example.

A way of conducting HIL simulation is to give inputs to the actual sensors of the systems instead of inputting them directly to the software [S6], [S72], [S83]. In the top example [S72], pictures and videos of 3D-printed stand-ins for microrobots are used to validate the depth estimation of optical tweezers. The images, however, are not fed to the software directly but through a microscope used in the system setup. Thus, the images and videos simulate the actual inputs that would be microrobots observed with the microscope. For a more imaginative example, in [S83], a physical steering wheel reacts to commands resulting from inputs in a traffic simulator, and the steering wheel angle is used as the metric for successful lane-keeping.

HIL is in many ways comparable to fully virtual simulation. The benefits and drawbacks of HIL are similar to those of fully virtual simulation. However, in comparison to a fully virtual simulation, faults can be injected into hardware and software in HIL simulations. Also, hybrid-faults originating from the combination of hardware and software can be monitored more easily [S6]. As hardware is included in the validation environment, not everything must be modelled in the simulator. However, HIL simulation is impractical or even impossible for most simple model-centred systems, as some physical components are required.

\textbf{System-in-the-loop simulation (SIL).}
In SIL simulation, the full system is placed into an artificial environment [S8], [S10], [S17], [S19], [S28], [S42], [S45], [S48], [S68]. Thus, the full system is under test. The environment can be a virtual or a physical one. In SIL simulation, the environment is clearly artificial and does not necessarily aim to replicate a real environment as is but some of its features. The only paper with a rigour score of 2.5 or higher is [S8]. Six out of nine systems were robots or cars.

SIL simulation can be conducted by picking presumed key tasks and challenges the system is expected to face and building an environment for simulating these [S8], [S68]. For example, in [S8], a service robot for an elderly care centre was initially validated in a laboratory setting with specific tasks before validating it in its real deployment environment. It was instructed to enter a room, to locate and approach a person, and to recognize the said person, etc. Thus, the setting does not replicate the environment, i.e., an elderly care centre, its obstacles, or its noise, but the system can be sent on similar missions than in a real environment. In other words, the robot does what it is supposed to do, but not in its final environment. In a similar manner, for the mining industry, a “simulated test environment” was built to replicate the difficulties a robot would face in a mine stope and to make the observations of the robot under test easier than in a real stope [S68].

The environment can also be virtual [S42]. Unlike in fully virtual simulation, the full system is used, not just its software. Essentially, virtual inputs are given to the full system and the system's behaviour is observed in real time. For example, in [S42], a UAV is tasked with following a virtual cow in a virtual environment while the vehicle itself is hovering in a laboratory. Instead of the UAV being modelled in the simulator, its actions are copied into it.

An interesting case of SIL simulation is presented in [S17]. For validating autonomous cars, the paper presents a mixed-reality approach, in which the avatars of real-world traffic participants are projected into a virtual simulator. These avatars can be projected into shared situations, which could potentially be dangerous in the real world. As a result, more realistic behaviour of the system under test and other traffic participants can be provided for the simulator compared with other, more modelling-oriented simulations.

The benefits of SIL simulation include safety and precision. As with other simulations, tests can be conducted even in scenarios that could be dangerous to the system or to other participants in real environments [S17], [S42]. On the one hand, having the full system in the validation process yields a more precise picture of how the system would behave in the validation scenarios. On the other hand, requiring real-world participants is heavy for the process compared with other simulation methods, and careless design in the environment may give a false impression of the system's functionality. This can be seen in our example paper [S8], in which the robot's interaction suggestion “touch me” resulted in people touching the robot anywhere instead of on its UI screen. This only occurred during later validations and did not come up during the simulation.

SIL simulation may be used as initial validation before other validation methods, as in [S8]. Success in a more controlled environment can provide confidence to update the system to even more realistic environments, for example, for a trial in the actual deployment environment.

\subsubsection{Trial} \label{trial}

This section describes the different trial variations. In the trials, the system, as is, is deployed and monitored in the final deployment environment or something close to it. The setting and goals are similar to alpha or beta testing in software engineering \cite{bourque2014guide}. However, while alpha and beta testing may be somewhat uncontrolled, the trials should be planned so that at least some key scenarios are covered. Additionally, compared with alpha and beta testing, the inclusion of potential end-users in the trials may not be necessary.

\textbf{Trial in a real environment.}
Trial in a real environment aims to validate the system by using the system as it would be used in the final deployment environment [S3], [S4], [S7], [S8], [S14], [S18], [S20], [S21], [S24], [S26], [S29], [S30], [S31], [S33], [S35], [S36], [S38], [S48], [S50], [S52], [S54], [S58], [S62], [S63], [S65], [S67], [S71], [S74], [S75], [S81], [S82], [S83], [S84], [S85], [S88], [S90]. Thus, the system is put under similar pressure as it is designed to be in during real use. The validation environment is a use case or an actual intended usage where it is observed, possibly accompanied with additional precautions. Also, ML systems that gather test data the way the complete system would be used fall into this category (e.g., [S24]). Papers with a high rigour score in QA are [S7], [S8], [S30], [S54], and [S71].

As an example, in [S7], a gaming-based application is built to assist in the screening of attention-deficit/hyperactivity disorder (ADHD). Test subjects are monitored while they play, and the data gathered by the instructor and the game are fed to the application, which gives a prediction of whether or not the subject has ADHD. This is then compared with the medical diagnoses of the subjects. The subjects are chosen from a larger sample of children, diagnosed with or without ADHD. This is to ensure that both key scenarios---the subject having and not having ADHD---are included in the validation process. Thus, even if the system is used in an actual use case, the validation settings can be arranged in a way that ensures that all the key scenarios are covered.

Because all papers using multiple validation methods also used trials, one approach for validating systems is to acquire a higher level of certainty in key scenarios by validating the system with other validation methods---such as  SIL simulation---first, and only then proceeding to the trial in a real environment (e.g., [S8]). Alternatively, trials can be conducted sequentially, starting from small and simple missions and then proceeding to the full scope [S26].

Trials in real environments hold one feature above other validation methods: the entire system is under test as intended. Thus, possible problems in component collaboration, communication, sensors, or usability become more easily evident, which may not be the case in simulations or less rigorous trials [S35].

On the other hand, rigorous trials are laborious. Even in the above example [S7], the number of test subjects is too small to give full confidence in the system. To their credit, however, the authors do discuss this when considering the validity of their application.

\textbf{Trial in a mock environment.} \label{mockTrial}
Trial in a mock environment is a form of trial in which the full system is validated in an environment that replicates an actual environment [S2], [S11], [S13], [S15], [S16], [S22], [S23], [S25], [S27], [S32], [S40], [S43], [S47], [S59], [S60], [S66], [S77], [S80], [S86]. There are no artificial components in the system under validation, but the environment is not real in a sense that it is designed to replicate or duplicate the actual deployment environment. The only paper in this category with a high rigour score is [S25].

In [S25], the trial of an intelligent music selection application was conducted in a laboratory setting. The application was installed on a phone, and it monitored the subjects' heartbeats while they were studying, trying to select music that would help the subjects be more efficient. Thus, the system was used as intended, but the environment was stripped of the noise of a real environment: the students were asked to study in an environment probably strange to them, and not necessarily when they were prepared to study. A trial can also be conducted in an environment that is a straight or partial copy of the deployment environment [S2], [S11]. It can also be a simple example part of the deployment environment [S15], [S16].

Malfunction in a mock environment instead of a real one relieves the possibility of major damage to the real environment or tasks at hand. Also, the system does not have to be brought into potentially dangerous environments, where it could be a distraction to other participants (e.g., [S27]).

The narrowed simplified environment can also serve as a downside. For example, [S25] discusses that people's emotions are influenced by many things. The authors do not elaborate, but one interpretation is that the limited test environment narrows down the validity of their emotion-centred application.

In some cases, distinguishing between a trial in a mock environment and SIL simulation (presented earlier in Section~\ref{simulation}) may be difficult or even arbitrary. However, the main difference is the environment, which is clearly artificial in SIL simulation  and does not necessarily aim to replicate the real environment as is but some of its features.

\subsubsection{Model-centred validation}

Model-centred validation is a set of validation methods that focus on ensuring the correct functioning of the used AI model [S12], [S46], [S49], [S51], [S55], [S57], [S61], [S64], [S69], [S76], [S79], [S82], [S85], [S87], [S89]. This type of validation is often data-centred and can be seen as having trust in the system by ensuring that the model works. These methods are mostly used in testing simple, model-centred systems (10 out of 15 of the systems are model-centred), but there are some more complicated systems, such as in [S69], which also rely on model-centred methods. No paper received a rigour score of 2.5 or higher. [S57] is used as the main example.

The approaches are often the same as approaches to ML model validation in general \cite{zhang2020machine}. Cross-validation is widely used: for example, in [S57], a subset of the training data is placed aside and later used to show that the system under test correctly predicts most of the test inputs. Another approach (referenced as “torture tests” in [S46]) tests the robustness of the model by intentionally altering or deleting small pieces of the model between inputs or by giving slightly modified inputs to see how this affects the model's performance. As such, model-centred validation includes ML model validation, as described in Section~\ref{sec:back&motiv}. Compared with the trials described in Section~\ref{trial}, it is noteworthy that in model-centred validation methods the data are not gathered the way a finalized system would.

Model-centred validation can be used as an initial validation before other validation methods [S82], [S85]. For example,  the model-centred system in [S82] is first tested statistically to investigate the---as they put it---model errors. Afterward, the system is further validated by comparing it with corresponding benchmarks and through a trial.

\subsubsection{Expert opinion} \label{expOpinion}

Expert opinion [S39], [S41], [S82] is a rarely reported method for validation. Its main feature is that the expertise required to conduct the validation is not present in the team responsible for the system but is obtained from the outside: the focus is less on rigorous testing and more on the expert's ideas of what is important in the system. The said expert is not necessarily an expert of software engineering or AI but rather an expert of the application domain, such as an intended end-user. Thus, expert opinion can be seen as a related term to beta testing of the system \cite{bourque2014guide}. No paper received a rigour score of 2.5 or higher. [S39] is used as the main example.

In [S39], a decision support system for emergency resource management was validated through interviews with fire officers. Example scenarios were presented to the experts, who then elaborated on which features they believe would add value to their work, and which deficits or delays are bearable and which are not. Also, the fire officers suggested that lower system accuracy is acceptable, thus suggesting thresholds for further development. A benefit of expert opinion is that the perceived usability and severity of potential problems are evaluated by those who actually face them in actual use.

Another approach is to compare the system's decisions with those of an expert or benchmark application [S41], [S82]. In this approach, the expert---or in the case of a benchmark application, “the expert application”---can provide insight into which scenarios---and failures---are crucial ones.

All three papers include the usage of the system by or presented to the expert. However, we decided to differentiate the category from trials, as other means are included, such as interviews. These include assessing the success of modelling and even the acceptable thresholds to be used in further validations [S39].

Compared with more rigorous trials, a short and rare session with experts may leave some problems related to key scenarios hidden. The same goes for benchmark applications, as comparisons made directly with them may not reveal what the benchmark application is not suitable for. Using expert opinions in addition to other validation methods is a way of tackling this issue [S39], [S82].

This kind of validation is most likely prone to the same problems as trials. With complex systems, presenting the system to an acceptable degree may turn out to be time-consuming. Also, as the expert comes from the end-user side and not necessarily from the production team, scheduling the validation sessions could prove problematic. As a benefit, more immature systems may possibly receive valuable feedback on how---and whether---to proceed with development.

\subsection{Descriptive analysis of continuous validation methods (RQ2)} \label{sec:analysisContVal}

As discussed at the end of Section~\ref{sec:back&motiv}, we report on system monitoring and other safety assurance as continuous validation methods. In this section, we describe these methods in more detail. We go over the basic principles of the different continuous validation methods (cf. Table~\ref{tab:contVal}). As less research has been conducted on these methods, we also synthesize the major benefits and drawbacks to a lesser degree than in the previous section. This section, along with Sections~\ref{sec:valStat}~and~\ref{sec:analysisVal}, answers RQ2.

\subsubsection{Failure monitoring}

Failure monitoring is a form of continuous validation for ensuring that system malfunctions do not go unnoticed during a system's life cycle. Failure monitors are designed to observe the system, its subsystems, and components and alert the user in case of malfunction or suspicious behaviour [S5], [S26], [S29], [S45]. No paper received a rigour score of 2.5 or higher. [S5] is used as the main example.

In [S5], a failure monitor of an autonomous vehicle observes other components. In case one shows signs of failure, the failure monitor alerts the human driver of a potential problem. The details of the observations, however, are not discussed in the paper.

Failure monitors only provide users with information and take no actions by themselves. Thus, human interference is always required in the case of a malfunction. This may be problematic or even dangerous in situations requiring rapid actions.

\subsubsection{Safety channel}

A safety channel is a backup component that takes over if primary components are compromised [S6], [S35], [S48], [S67], [S78]. This way the systems can continue to function in the presence of danger or malfunction, i.e. they may “fail safely”, as [S67] puts it. Safety channels are a form of continuous validation. [S78] is the only paper with a rigour score of 2.5 or higher.

In [S78], if the sensor of an autonomous car detects another vehicle too close in front, the safety channel forces the car to slow down. Thus, the system's safety constraints are compromised, and the main controller's directions are disregarded. Similarly, in [S48], if physical obstacles are recognized, the robot's functionalities are reduced accordingly to ensure the safety of the situation.

A safety channel may be a simpler, yet more robust version of the main controller of a car [S6]. Alternatively, in a group of systems, other systems can take over the tasks of a failed system [S35]. 

As the safety channel's primary objective is to keep the system running safely when needed, it usually reduces system functionality. This could mean that it cannot necessarily keep the system safe for longer periods and thus requires interference at some point.

\subsubsection{Redundancy}

In redundancy, critical components are duplicated within the system [S15]. This way the duplicating components can work as a backup along the lines of a safety channel. While a safety channel takes over in case of a malfunction or safety hazard and attempts to keep the situation safe, redundant components take over the malfunctioning component they duplicate and attempt to keep up initial functionality. In addition, the components can share the workload of the primary components if the performance level of the system drops. [S15] has a rigour score of 2.

In [S15], every component responsible for a task in the software controller of an autonomous car is replicated to achieve redundancy. Implementation details are not really described in the paper, but when introducing the idea of using redundancy, they cite Jiang \& Yu \cite{jiang2012fault}, on which we base much of this description.

Redundancy makes systems more tolerant to faults but also adds costs [S15]. This is due to the need to multiply the same components. In addition, the state space of the system grows, which may make it more difficult to monitor.

\subsubsection{Voting}

Various intelligent components can be implemented in a system to execute the same task and then vote among each other on the action to be taken [S20]. Thus, damage caused by a single failing component is reduced if the others are still functional. Voting is a form of continuous validation. [S20] has a rigour score of 2.

In [S20], a robot assembles small objects by mimicking a person performing the same task. Voting is involved when determining which motions the person takes based on the video input, and hence, what the robot should do. Unfortunately, further details of the voting process itself are not clearly discussed in the paper.

Similar to redundancy, voting requires multiple components capable of performing the same task and plan to be voted on. This raises the costs of the system.

\subsubsection{Output and input restrictions}

Output restrictions are hard limits given to a system [S50], [S78]. This is to continuously validate that the system will not take actions that are known to be potentially dangerous or false. This way the system can also work more predictably. [S78] has a rigour score of 2.5.

In [S78], an autonomous car has limited outputs if certain criteria apply. For example, if another vehicle is directly adjacent to the vehicle-under-test, the vehicle-under-test will not change into that lane, no matter what the controller suggests.

Input restrictions, in turn, limit what kinds of situations the system is allowed to handle autonomously [S74], [S87]. By limiting inputs,  the system can be ensured to not attempt to handle a situation it is not designed or trained for. Neither of the papers received a rigour score of 2.5 or higher. [S87] is used as an example.

In [S87], a report assignment tool is built to automatically assign incoming reports to the corresponding development team. While making predictions, the system does not take into account certain strings (such as “error” and “ericsson”) in the input, as the commonality of these strings in the product reports decreases prediction accuracy.

\bigskip
\bigskip
\bigskip
\bigskip\bigskip\bigskip

\section{Discussion}

\begin{table*}[t]
    \caption{Summary of validation methods (RQ2).}
    \centering

\begin{tabularx}{\textwidth}{p{4.7cm} X p{1cm} } 

Validation method & Description & Exemplar \\
\hline
\emph{Validation:}\\
Simulation:\\
Fully virtual simulation&
The  deployment environment of the system is replicated with a virtual simulator& [S78] \\

Hardware-in-the-loop simulation &
A virtual simulator that contains also some non-virtual components&
[S72]\\

System-in-the-loop simulation&
The system in an artificial environment&
[S8]\\
[3mm] 
Trial:\\
Trial in a real environment&
The system is used as it would be used in the final deployment environment&
[S7]\\

Trial in a mock environment&
The  system is used in an environment that replicates an actual environment&
[S25]\\
[3mm] 

Model-centered validation&
Validation focusing solely on validating the model&
[S57]
\\
[3mm] 
Expert opinion&
The system is assessed against expert's opinion&
[S39]
\\ \\

\emph{Continuous validation:}\\
Failure monitor&
System's malfunction detection&
[S5]\\

Safety channel&
Backup component that takes over if the primary components are compromised&
[S78]\\

Redundancy&
Critical components are duplicated in the system&
[S15]\\
Voting& 
Different components perform the same task and  vote on the action to be taken &
[S20]\\
Output and input restrictions&
Hard limits given to the system input and output&
[S78], [S87]
\\

\hline
\end{tabularx}
\label{tab:summary}
\end{table*}


\subsection{Overview of results}

Practical intelligent systems are receiving rising interest the research literature. We found 90 primary studies in total. The number of publications in the area has risen for five consecutive years. The sudden peak in numbers from 2012 to 2013, however, remains a mystery, as the papers are published on several forums.

The quality assessment of the papers revealed a wide disregard for validity discussion among the papers. While more than half of the papers received a high score in context, design, or both, the vast majority of the papers in no way mentioned the validity of their research. Explicitly discussing validity threats or limitations of the research is important, especially because the majority of research is conducted in academic, simplified contexts. Another significant observation is the lack of realistic industrial studies, i.e. studies in evidence levels 5 and 6, which would increase validity and facilitate research on continuous validation.

Our focus was only on papers with solid empirical evidence representing at least level L4 in the taxonomy of Alves et al. \cite{alves2010requirements}. Thus, the numbers do not represent the research literature as a whole. 
Thus, the selection of papers aims to cover more mature research and disregard immature solution proposals that may become breakthroughs but equally could just fade away.
Even the original SLR guidelines~\cite{kitchenham2007guidelines} emphasize empirical evidence, and recent experiences concerning systematic reviews \cite{raatikainen2019software} suggest limiting analysis to papers with a higher level of evidence and quality, which grants a better stand when analysing the papers qualitatively.

\subsection{Answer to RQ1: Characteristics of validated systems}

As for RQ1, a wide variety of systems has been reported in the research literature, as we identified 13 classes of systems. Considering this variety over different aspects of the systems (detailed below), it is difficult to pinpoint large common characteristics beyond the apparent dominance of ML (at least 87\% of all included studies). Thus, it may be that the practical use of AI is not naturally characterizable in regards to the specific domains, complexity, or malfunction impact but broadly spreads across various dimensions.

The variety can be seen in the large numbers of application domains. Large numbers in certain domains may be a result of our search method, but the diversity of the identified domains indicates wide applicability of AI research interests. Of course, the actual variety was even vaster, as, for example, commercial and industrial domains contain systems very different from one another but that were “forced” into the same category. 

The same variety goes for tasks performed by systems. Recognition \& classification are by far the most common tasks assigned to systems with AI, but they are by no means the only ones. Variety can be seen as an implication of the flexibility of AI techniques. However, the popularity of a few tasks may pose a threat where progress is not made holistically but research rather focuses on minor, local improvements in known, well-suited tasks, such as recognition and classification, and applies the task to similar problems in slightly different domains.

Variety can also be found in the impact of system malfunction. The impact of the systems ranges from nuisance to lethal. This supports the previous implications that AI raises interest in a wide range of problems, whether they are dangerous or not. However, systems concerning people's privacy or the risk of discrimination are scarce and far apart. Also, every system may not have to be validated to the same degree: a system annoying someone is much more forgivable than a system that kills someone. Thus, the impact level of the system malfunction should be taken into account when building a system.

However, while the systems are widely spread across different categories, they are heavily characterized by narrow application within these categories. Many categories in each aspect were represented by only a few or even only one system. For example, some domains were significantly less represented than others, as well as less consistently represented. Robotics overshadowed agriculture and even medical applications, and this difference has been growing in recent years. Recognition \& classification dominated the system tasks. This may be due to researchers' accessibility to certain resources, such as robots, or to the sheer popularity of certain applications, such as autonomous cars, or perhaps to the maturing technology's ability to finally be applied in these more complex domains. Nonetheless, the less popular domains should not go unnoticed in further studies, as lower popularity does not mean they are less valuable, and AI could prove to be a great asset in problems such as crop ripeness assessment and crop disease detection, or MRI scan assistance. This is especially noteworthy considering that practical research on these potentially hugely benefitting areas could help the technologies reach the practitioners and should be taken to account in research. This has not always been the strong point of software engineering research \cite{sjoberg2007future, Ivarsson2011}.

We also note that a large portion of the tasks assigned to the systems are relatively low level in complexity. For example, the largest category 'recognition \& classification' is not an easy task, but it is often relatively simple in terms of the problems that it solves: Is there mould in the goods? Are these pictures taken of the same person? Is the crop ripe for harvesting? Now compare this to the rare task of assembly, which can be seen as consisting of recognition \textit{and} something more: recognize the relevant pieces and put them together in a purposeful manner. Thus, it seems that only a little of the research on practical AI systems is focused on highly complex tasks. This may indicate that the time is only just ripening for the field to take on these complex problems in practical settings.

\subsection{Answer to RQ2: Validation methods}

As for the second research question (RQ2, see summary in Table~\ref{tab:summary}), a taxonomy of validation methods was extracted from the papers (see also Figure~\ref{fig:taxonomy}). The validation methods were identified, described, and provided with exemplars from the papers. The taxonomy consists of four main validation methods: simulation, trial, model-centred validation, and expert opinion. The simulation is further divided into fully virtual simulations, hardware-in-the-loop simulations, and system-in-the-loop simulations. The trial is further split into trials in a real environment and trials in a mock environment. In 77 papers, a single validation method was reported. This means that only 13 papers reported using more than one method.

Trial was the most used validation method among the included papers and a popular approach across domains. This suggests that researchers prefer bringing their systems close to their intended settings. Trials can be cheaper in situations where the final environment is simple enough, as building a high-quality simulation is likely to add costs. Also, trials are not prone to deficits in simulators or data, which speaks for using trials: if a system works, it works because it should and not because of imprecise modelling. Considering the close relation of trials to alpha and beta testing and their established position in software engineering, it may not be that far-fetched to assume that trials are also commonly practiced in the industry.

However, even otherwise well-constructed papers, such as~[S7], rely on very small sample sizes in their trials. This could be because, as declared by \cite{kalra2016driving, koopman2016challenges}, some systems are laborious, hard, or infeasible to test by trial. Additionally, some systems may be considered too hazardous to be validated by trials, such as safety-critical systems. These seem to be reflected in the numbers: a large portion of more complex cyber-physical systems, such as cars and robots, rely on simulation over trials. This can be seen in the rising interest in simulations along with the rising interest in these domains. Thus, it seems that researchers believe that high-level simulations are the way to tackle these problems pertaining to trials. However, this does not come without its problems, as stated in [S1]: even the slightest deviations between a simulation and the real environment can have dire consequences, and according to [S78], the simulator's validity is hard to prove. In addition, many---especially virtual---simulators were presented with one or two example cases. This may suggest that current high-level simulators have been difficult to find, as suggested by Gao et al. \cite{gao2019ai}, or, for some reason, they have not been satisfactory to the researchers at the time of developing their system.

Model-centred validation is also widely applied, especially among model-centred systems. Its rarity in validating other kinds of systems can be considered a relief when considering the earlier idea that model performance may not be the best metric for validating systems \cite{nair2017using}.

Expert opinion is rarely reported in the research literature. We also note that we do not know how common it is for researchers to not validate their systems at all, as a presented validation method was an inclusion criterion in our paper selection phase.

As data-driven ML testing in general, a large part of the validations based on model-centred approaches are prone to data problems \cite{zhang2020machine}. However, if enough good-quality data are available, model-centred methods are an efficient and potentially faster way of validating model-centred systems or at least models used in the systems. If a system consisted of multiple components, we would advise  to  carefully consider whether or not the model-centred approach is a suitable validation method.

Many traditional software testing methods (e.g. unit, integration, or regression testing) were not mentioned in the papers. This could be due to the Oracle problem and the differing requirements of traditional and AI systems \cite{vassev2014autonomy}: testing non-deterministic systems is difficult with deterministic tests. Usually in traditional software testing, the indication of correct functioning is that with a given series of inputs, a certain state or output is reached at any given time. But how to choose the deterministic tests that indicate that a possibly non-deterministic system works as intended? It may well be that getting a different outcome at two points in time \emph{is} the intended or acceptable functionality. Respectively, what to actually test with unit tests and what the actual desired outcome is can be difficult to identify. For example, when an autonomous car is sent on a mission, it goes through an enormous set of inputs, most of which have multiple outputs that can be considered correct, as well as multiple outputs that can be considered incorrect, with possibly no indication of which can be considered the best. How do you choose which ones to implement? However, this does not mean that no such tests were used in the systems at all, as most systems do not consist solely of non-deterministic components. We even consider a total lack of these tests unlikely. Nevertheless, even if the tests are there, the researchers have not considered them essential enough to be mentioned when validating their entire systems. This, too, could underline the troubled relationship between traditional testing to non-determinism. That being said, traditional testing may also have a place in validating AI systems. Not every AI system changes its behaviour autonomously. Also, those that do adapt over time can have some corner cases in which functionality should remain the same, no matter how much they should adapt to their environment. For example, if the sensors of a robot indicate a steep slope in front of it, the robot should probably always stop, and a financial bot should probably never grant loans to someone with no income or assets. Such corner cases could potentially be used in regression testing, at least in self-learning systems, to indicate possible deterioration of the system over time. This may be an interesting topic for future research.

Only 14 papers reported using a method for continuous validation. Failure monitoring, safety channel, redundancy, voting, and output and input restrictions were identified as methods for continuous validation (Table~\ref{tab:summary}). The number of studies carrying out continuous validation seems low considering that other literature has emphasized the importance of monitoring the system even if the initial validation was successful \cite{breck2017ml, zhang2020machine}. The methods were often described cursorily, with little emphasis, and often in papers with low QA scores. However, we must note that not mentioning a method does not necessarily mean that one does not exist or was not intended: many of the papers describe a system to the point of its initial validation, and the settings are typically academic rather than industrial use. Thus, methods for continuous validation may not have been in the scope of the paper and were left out. Also, 13 of the papers mainly focused on describing a tool or method for initial validation, leaving little room for describing the validated system and its life cycle altogether. Either way, research reporting on continuous validation is far from desired.

Overall, initial validation seems to be emphasized in the literature over continuous validation. This may not be surprising, as this is the traditional way of doing things: to put it simply, when the system seems ready to be deployed, it is tested and then it is ready. However, considering again the differing requirements of AI and traditional systems \cite{vassev2014autonomy}, this may not be the case. If we once again return to the example of a self-configuring system, the initial validation would probably include the validation of self-configurability and the initial configuration. However, should the system reconfigure itself during deployment, the new configuration should probably be validated as well, to ensure the system still meets its requirements. The same goes, let us say, for the recommendation algorithm of a video streaming platform, the user base of which may constantly be changing: with no continuous validation, the algorithm may fail to meet its requirements, unbeknownst to the developers and users.

Our validation and continuous validation classifications were based on and emerged from the primary studies. As such, it is not our aim to assess the completeness of the classifications in this study. As various taxonomies may also be useful, a future challenge is to assess, extend, and refine these taxonomies for validity and utility for classifying and understanding AI validation.

\section{Threats to validity} \label{sec:threat}

The numbers presented in this paper only investigate papers published over the years by the scientific community. Thus, they can give an indication of what has been of interest to researchers in the AI field. However, they do not necessarily reflect what is happening in the industry, nor should they be considered representative of the commonness or rareness of various systems or validation methods used in the industry.

Our choice of an automatic search may have left some relevant papers out, as the search is so dependent on the successful compaction of the search string. This dependability can probably be seen in the high occurrence rates of the domains that were specifically mentioned in the search string. Although the use of certain domains may have influenced the included domains, we did not intentionally select the domains in favour of something. Instead, we selected all the domains we identified to be relevant to the topic but that could be at risk of being excluded if more general terms were used solely. The use of trendy domains may also be one of the reasons why recent papers were much more common than older ones. However, the wide variety of domains, tasks, and other characteristics, along with the number of papers found is encouraging when considering the trustworthiness and coverage of the search.

The dominance of ML in the results poses a threat to the generalizability of the results. As most papers presented systems utilizing ML, the results would be safest to interpret as results concerning ML literature, not AI in general. The skewed results could be due to  the search terms 'ML' and 'validation' usually going hand-in-hand. Also, it is not uncommon for technologies to "lose" their status as AI, and not be referred to as AI anymore, thus leaving the stage for ML to roam free. Finally, the amount and rigour of empirical research in computer science and software engineering have risen over the years. Older research concerning AI, which is more often theoretical or basic research, and predates the resurgence of ML, may not have been conducted or reported in the required manner to reach the desired level of relevance.

No description of validation methods was based on a single paper, but in the case of continuous validation, the evidence was scarce. This gives confidence in the construct validity of the descriptions, as commonalities could be used as the core of each description. Thus, some light is shed on the validation of entire AI systems, as requested by Zhang et al. \cite{zhang2020machine}. Methods for continuous validation, however, were often described in less detail and in fewer papers. Also, the analysis was largely done by a single researcher, which leaves the taxonomy prone to bias.

Bias is of course a potential threat in other forms as well. One potential source of bias is caused by most of our paper selection and analysis being performed by the first author. Unclear cases were evaluated with other researchers, but there is no guarantee that what the first author views as a clear case would also be clear and unambiguous to others. This applies to both paper selection and the analysis of the final set of papers. Some relevant papers may have been excluded based on a mere interpretation of them, and some analyses may not be accurate because of how they were interpreted. Bias may also affect the terms used in the search string. The terms may not be biased in and of themselves, but the choices of including or excluding a term are of course guided by what the authors consider to be intelligent behaviour and what type of research the authors are aware of in the first place. The latter is especially true in the case of including domains thought to be relevant yet at risk of not being included when only using general terms. Other domains were considered, but nothing can guarantee that no choices were misguided, or that there are no domains that should have been included but were not considered for some reason.

Sometimes differentiating between validation methods---for example, system-in-the-loop simulation and trial in a mock environment---can be difficult or even arbitrary. This can be problematic when categorizing papers precisely. However, as no description of a validation method was based on a single paper, categorizing one or two papers “wrongly” should not pose any threat to the validity of the descriptions. More precise and consistent use of terminology in publications could ease further assessments.

\section{Conclusions}
We presented a systematic literature review on the validation methods of AI systems based on 90 primary studies. The primary studies represent 14 domains carrying out 18 tasks and their impact on malfunction ranges from nuisance to lethal, which demonstrates a wide variety and broad application of AI technologies. As our selection focused on studies with solid empirical evidence, these studies represent relatively mature practical applications rather than immature solution proposals in early-stage research. However, most of these systems applied ML.
Concerning the quality of the papers, most described their context and design well but lacked discussion on the validity of the study. This is something the research community should be more aware of.

We identified a taxonomy of four validation methods: simulation, trial, model-centred validation, and expert opinion. In addition to this, we described the validation methods and their common variations, provided with examples from empirical studies. Thus, the taxonomy should be easily used as a basis for further attempts to synthesize the validation of AI systems or even to propose general ideas on how to validate systems. However, as the taxonomy emerged from the included papers alone, it may not be complete. Thus, further assessment and refinement may be needed for it to be fully utilized in understanding the validation of AI systems.

A set of continuous validation methods was also presented, consisting of seven classes. These methods are  described in only 13 papers and in less detail. Thus, not only is the set in need of assessment, but we also suggest placing more emphasis on these methods in system descriptions, as they can be vital for the safety and functionality of systems during their life cycles.

Given our results, it seems that more care needs to be taken, specifically for a few concerns. Overall, study validity needs to be taken more into account in the measures and discussions, as many papers deal with, e.g., academic settings and small samples, yet do not discuss this in the paper. In terms of research design, on the one hand, academic research settings could improve their validity by more commonly adopting multiple validation methods to mitigate the threats to study validity. On the other hand,  more research, assessment, and development need to be conducted in realistic, industrial settings. This, in particular, would also allow to better adopt and study continuous validation methods, as they are currently discussed in very little detail. More research should be done on continuous validation if these ideas are to be generalized. Also, if high-quality simulators are indeed intended to carry out the validation of high-complexity systems in the future, more research is to be conducted on them. Many simulators in the papers seem small in scale, and, as discussed by the papers, their validity is difficult to prove. Initiatives e.g., Simulink and Dronology have taken fine first steps in the area but should be accompanied by others, possibly aimed at different domains. Finally, we would also like to see more research on the rarer application domains, along with completely new ones.

\section*{Acknowledgement}
This work was labelled by ITEA3 and funded by local authorities (Business Finland) under grant agreement “ITEA-2019-18022-IVVES” https://ivves.weebly.com/.
\clearpage

\onecolumn																											
\tablefirsthead{ID & Author(s) (F1) & Year (F3) & Title (F2) and DOI  \\ \hline}																											
\tablehead{\multicolumn{4}{l}{\footnotesize\sl Table \ref{tab:PrimaryStudies} continued}\\ ID  & Author(s) (F1) & Year (F3) & Title (F2) and DOI  \\ \hline}																											
\tablecaption{The bibliographic data of the primary studies. The columns are the fields of the data extraction form (Section \ref{sec:InformationExtraction}).}																											
\footnotesize																											
\begin{supertabular}{p{0.2cm}p{6cm}p{1.2cm}p{9cm}}

S1	&	Li, Y.; Tao, J.  \&  Wotawa, F.	&	2020	&	Ontology-based test generation for automated and autonomous driving functions	(10.1016/j.infsof.2019.106200)	\\																			
S2	&	Vithanage, R.; Harrison, C.  \&  De Silva, A.	&	2019	&	Autonomous rolling-stock coupler inspection using industrial robots	(10.1016/j.rcim.2019.03.009)	\\																			
S3	&	Mannini, A.  \&  Intille, S.	&	2019	&	Classifier Personalization for Activity Recognition Using Wrist Accelerometers	(10.1109/jbhi.2018.2869779)	\\																			
S4	&	Lu, B.; Wang, L.; Liu, J.; Zhou, W.; Guo, L.; Jeong, M.-H.; Wang, S.  \&  Han, G.	&	2019	&	LaSa: Location Aware Wireless Security Access Control for IoT Systems	(10.1007/s11036-018-1088-x)	\\																			
S5	&	Jha, S.; Banerjee, S.; Tsai, T.; Hari, S.; Sullivan, M.; Kalbarczyk, Z.; Keckler, S.  \&  Iyer, R.	&	2019	&	ML-Based Fault Injection for Autonomous Vehicles: A Case for Bayesian Fault Injection	(10.1109/dsn.2019.00025)	\\																			
S6	&	Fu, Y.; Terechko, A.; Bijlsma, T.; Cuijners, P.; Redegeld, J.  \&  Ors, A.	&	2019	&	A Retargetable Fault Injection Framework for Safety Validation of Autonomous Vehicles	(10.1109/icsa-c.2019.00020)	\\																			
S7	&	Mwamba, H.; Fourie, P.  \&  van den Heever, D.	&	2019	&	PANDAS: Paediatric attention-deficit/hyperactivity disorder application software	(10.3390/app9081645)	\\																			
S8	&	Portugal, D.; Alvito, P.; Christodoulou, E.; Samaras, G.  \&  Dias, J.	&	2019	&	A Study on the Deployment of a Service Robot in an Elderly Care Center	(10.1007/s12369-018-0492-5)	\\																			
S9	&	Domínguez, C., Martínez, J., Busquets-Mataix, J.V. et al	&	2019	&	Human–computer cooperation platform for developing real-time robotic applications.	(10.1007/s11227-018-2343-4)	\\																			
S10	&	Ramezani, S.  \&  Hasanzadeh, R.	&	2019	&	Defect detection in metallic structures through AMR C-scan images using deep learning method	(10.1109/pria.2019.8786029)	\\																			
S11	&	Lin, W.; Anwar, A.; Li, Z.; Tong, M.; Qiu, J.  \&  Gao, H.	&	2019	&	Recognition and Pose Estimation of Auto Parts for an Autonomous Spray Painting Robot	(10.1109/tii.2018.2882446)	\\																			
S12	&	Kachamas, P.; Akkaradamrongrat, S.; Sinthupinyo, S.  \&  Chandrachai, A.	&	2019	&	Application of artificial intelligent in the prediction of consumer behavior from facebook posts analysis	(10.18178/ijmlc.2019.9.1.770)	\\																			
S13	&	Li, T.-H.; Kuo, P.-H.; Tsai, T.-N.  \&  Luan, P.-C.	&	2019	&	CNN and LSTM Based Facial Expression Analysis Model for a Humanoid Robot	(10.1109/access.2019.2928364)	\\																			
S14	&	Cai, C.-H.; Sun, J.  \&  Dobbie, G.	&	2018	&	B-Repair: Repairing B-models using machine learning	(10.1109/iceccs2018.2018.00012)	\\																			
S15	&	Yan, R.; Yang, J.; Zhu, D.  \&  Huang, K.	&	2018	&	Design verification and validation for reliable safety-critical autonomous control systems	(10.1109/iceccs2018.2018.00026)	\\																			
S16	&	Ericsson, M.; Zhang, X.  \&  Christiansson, A.-K.	&	2018	&	Virtual Commissioning of Machine Vision Applications in Aero Engine Manufacturing	(10.1109/icarcv.2018.8581207)	\\																			
S17	&	M. R. Zofka et al.	&	2018	&	Traffic Participants in the Loop: A Mixed Reality-Based Interaction Testbed for the Verification and Validation of Autonomous Vehicles	(10.1109/itsc.2018.8569226)	\\																			
S18	&	Nishimi, T.; Sato, Y.; Kajihara, S.  \&  Nakamura, Y.	&	2018	&	Good die prediction modelling from limited test items	(10.1109/itc-asia.2018.00030)	\\																			
S19	&	Seweryn, K.; Rybus, T.; Colmenarejo, P.; Novelli, G.; Oles, J.; Pietras, M.; Sasiadek, J.; Scheper, M.  \&  Tarenko, K.	&	2018	&	Validation of the Robot Rendezvous and Grasping Manoeuvre Using Microgravity Simulators	(10.1109/icra.2018.8460475)	\\																			
S20	&	Wang, Y.; Xiong, R.; Yu, H.; Zhang, J.  \&  Liu, Y.	&	2018	&	Perception of Demonstration for Automatic Programing of Robotic Assembly: Framework, Algorithm, and Validation	(10.1109/tmech.2018.2799963)	\\																			
S21	&	Coronel-Reyes, J.; Ramirez-Morales, I.; Fernandez-Blanco, E.; Rivero, D.  \&  Pazos, A.	&	2018	&	Determination of egg storage time at room temperature using a low-cost NIR spectrometer and machine learning techniques	(10.1016/j.compag.2017.12.030)	\\																			
S22	&	Lyons, D.; Arkin, R.; Jiang, S.; O'Brien, M.; Tang, F.  \&  Tang, P.	&	2017	&	Performance verification for robot missions in uncertain environments	(10.1016/j.robot.2017.07.001)	\\																			
S23	&	Mattos, D.; Bosch, J.  \&  Olsson, H.	&	2017	&	Your system gets better every day you use it: Towards automated continuous experimentation	(10.1109/seaa.2017.15)	\\																			
S24	&	Cavallo, D.; Cefola, M.; Pace, B.; Logrieco, A.  \&  Attolico, G.	&	2017	&	Contactless and non-destructive chlorophyll content prediction by random forest regression: A case study on fresh-cut rocket leaves	(10.1016/j.compag.2017.06.012)	\\																			
S25	&	Chiu, M.-C.  \&  Ko, L.-W.	&	2017	&	Develop a personalized intelligent music selection system based on heart rate variability and machine learning	(10.1007/s11042-016-3860-x)	\\																			
S26	&	Balta, H.; Bedkowski, J.; Govindaraj, S.; Majek, K.; Musialik, P.; Serrano, D.; Alexis, K.; Siegwart, R.  \&  De Cubber, G.	&	2017	&	Integrated Data Management for a Fleet of Search-and-rescue Robots	(10.1002/rob.21651)	\\																			
S27	&	Shi, S.; Wu, H.; Song, Y.  \&  Handroos, H.	&	2017	&	Mechanical design and error prediction of a flexible manipulator system applied in nuclear fusion environment	(10.1108/ir-04-2017-0066)	\\																			
S28	&	Oleś, Jakub, et al.	&	2017	&	A 2D microgravity test bed for the validation of space robot control algorithms	(10.14313/jamris\_2-2017/21)	\\																			
S29	&	Fotinea, S.-E.; Dimou, A.-L.; Efthimiou, E.; Tzafestas, C.; Goulas, T.  \&  Pitsikalis, V.	&	2016	&	The MOBOT human-robot interaction-Showcasing assistive HRI	(10.1145/3003733.3003812)	\\																			
S30	&	Jonsson, L.; Borg, M.; Broman, D.; Sandahl, K.; Eldh, S.  \&  Runeson, P.	&	2016	&	Automated bug assignment: Ensemble-based machine learning in large scale industrial contexts	(10.1007/s10664-015-9401-9)	\\																			
S31	&	Xu, D.; Chen, Y.; Chen, X.; Xie, Y.; Yang, C.  \&  Gui, W.	&	2016	&	Multi-model soft measurement method of the froth layer thickness based on visual features	(10.1016/j.chemolab.2016.03.029)	\\																			
S32	&	Silva, R.; Fleury, A.; Martins, F.; Ponge-Ferreira, W.  \&  Trigo, F.	&	2015	&	Identification of the state-space dynamics of oil flames through computer vision and modal techniques	(10.1016/j.eswa.2014.10.030)	\\																			
S33	&	Navarro, J.; Parada, G.  \&  Duenas, J.	&	2014	&	System Failure Prediction through Rare-Events Elastic-Net Logistic Regression	(10.1109/aims.2014.19)	\\																			
S34	&	Aguilera, S.; Torres-Torriti, M.  \&  Auat, F.	&	2014	&	Modeling of skid-steer mobile manipulators using spatial vector algebra and experimental validation with a compact loader	(10.1109/iros.2014.6942776)	\\																			
S35	&	Portugal, D.  \&  Rocha, R.	&	2013	&	Distributed multi-robot patrol: A scalable and fault-tolerant framework	(10.1016/j.robot.2013.06.011)	\\																			
S36	&	Azadeh, A.; Ghaderi, S.; Anvari, M.; Izadbakhsh, H.; Rezaee, M.  \&  Raoofi, Z.	&	2013	&	An integrated decision support system for performance assessment and optimization of decision-making units	(10.1007/s00170-012-4387-6)	\\																			
S37	&	Serrano, E.  \&  Botia, J.	&	2013	&	Validating ambient intelligence based ubiquitous computing systems by means of artificial societies	(10.1016/j.ins.2010.11.012)	\\																			
S38	&	Ranaweera, K.; Ruwanpura, J.  \&  Fernando, S.	&	2013	&	Automated real-time monitoring system to measure shift production of tunnel construction projects?	(10.1061/(asce)cp.1943-5487.0000199)	\\																			
S39	&	Pottebaum, J.; Artikis, A.; Marterer, R.  \&  Paliouras, G.	&	2012	&	User-oriented evaluation of event-based decision support systems	(10.1109/ictai.2012.30)	\\																			
S40	&	Torta, E.; Cuijpers, R.; Juola, J.  \&  Van Der Pol, D.	&	2012	&	Modeling and testing proxemic behavior for humanoid robots	(10.1142/s0219843612500284)	\\																			
S41	&	Begum, S.; Ahmed, M.; Funk, P.  \&  Filla, R.	&	2012	&	Mental state monitoring system for the professional drivers based on Heart Rate Variability analysis and Case-Based Reasoning	(NA)	\\																			
S42	&	Chen I.YH., MacDonald B.  \&  Wünsche B.	&	2012	&	Evaluating the effectiveness of mixed reality simulations for developing UAV systems	(10.1007/978-3-642-34327-8\_35)	\\																			
S43	&	Faria, A. W., Menotti, D., Pappa, G. L., Lara, D. S.,  \&  Araújo, A. D. A.	&	2012	&	A methodology for photometric validation in vehicles visual interactive systems	(10.1016/j.eswa.2011.09.126)	\\																			
S44	&	Campuzano, F.; Serrano, E.  \&  Botia, J.	&	2012	&	Towards socio-chronobiological computational human models	(10.1007/978-3-642-34654-5\_40)	\\																			
S45	&	Wang, X.; Liang, B.; Li, C.  \&  Xu, W.	&	2008	&	The ground-based validation technology of teleoperation for space robot	(10.1109/ramech.2008.4681426)	\\																			
S46	&	Baumeister, J.; Bregenzer, J.  \&  Puppe, F.	&	2007	&	Gray box robustness testing of rule systems	(10.1007/978-3-540-69912-5\_26)	\\																			
S47	&	Lye, S.; Lee, S.  \&  Chew, B.	&	2004	&	Virtual design and testing of protective packaging buffers	(10.1016/j.compind.2003.01.001)	\\																			
S48	&	Raineri, M.; Perri, S.  \&  Lo Bianco, C. G.	&	2019	&	Safety and efficiency management in LGV operated warehouses	(10.1016/j.rcim.2018.11.003)	\\																			
S49	&	Li, C.; Kong, F.; Wang, K.; Xu, A.; Zhang, G.; Xu, N.; Liu, Z.; Guo, H.; Wang, X.; Liang, K.; Yuan, J.; Qi, S.  \&  Jiang, T.	&	2019	&	Microscopic Machine Vision Based Degradation Monitoring of Low-Voltage Electromagnetic Coil Insulation Using Ensemble Learning in a Membrane Computing Framework	(10.1109/ACCESS.2019.2928025)	\\																			
S50	&	Prado, A. J.; Michalek, M. M.  \&  Cheein, F. A.	&	2018	&	Machine-learning based approaches for self-tuning trajectory tracking controllers under terrain changes in repetitive tasks	(10.1016/j.engappai.2017.09.013)	\\																			
S51	&	Stamate, C.; Magoulas, G. D.; Kueppers, S.; Nomikou, E.; Daskalopoulos, I.; Luchini, M. U.; Moussouri, T.  \&  Roussos, G.	&	2017	&	Deep Learning Parkinson's from Smartphone Data	(10.1109/percom.2017.7917848)	\\																			
S52	&	Kolhe, S.; Kamal, R.; Saini, H. S.  \&  Gupta, G. K.	&	2011	&	A web-based intelligent disease-diagnosis system using a new fuzzy-logic based approach for drawing the inferences in crops	(10.1016/j.compag.2011.01.002)	\\																			
S53	&	Prasad, M.  \&  Mohan, S.	&	2005	&	Fuzzy logic model for multi-purpose multi-reservoir system	(10.1007/0-387-29295-0\_36)	\\																			
S54	&	Zheng, Y.; Xie, X.; Su, T.; Ma, L.; Hao, J.; Meng, Z.; Liu, Y.; Shen, R.; Chen, Y.  \&  Fan, C.	&	2019	&	Wuji: Automatic Online Combat Game Testing Using Evolutionary Deep Reinforcement Learning	(10.1109/ASE.2019.00077)	\\																			
S55	&	Pfahl, D.; Karus, S.  \&  Stavnycha, M.	&	2016	&	Improving Expert Prediction of Issue Resolution Time	(10.1145/2915970.2916004)	\\																			
S56	&	Balkan, A.; Tabuada, P.; Deshmukh, J. V.; Jin, X.  \&  Kapinski, J.	&	2017	&	Underminer: A Framework for Automatically Identifying Nonconverging Behaviors in Black-Box System Models	(10.1145/3122787)	\\																			
S57	&	Zong, L.	&	2019	&	Classification Based Software Defect Prediction Model for Finance Software System - An Industry Study	(10.1145/3374549.3374553)	\\																			
S58	&	Hu, G.; Zhu, L.  \&  Yang, J.	&	2018	&	AppFlow: Using Machine Learning to Synthesize Robust, Reusable UI Tests	(10.1145/3236024.3236055)	\\																			
S59	&	Paudyal, P.; Lee, J.; Banerjee, A.  \&  Gupta, S. K. S.	&	2019	&	A Comparison of Techniques for Sign Language Alphabet Recognition Using Armband Wearables	(10.1145/3150974)	\\																			
S60	&	Amsuess, S.; Goebel, P.; Graimann, B.  \&  Farina, D.	&	2015	&	 A Multi-Class Proportional Myocontrol Algorithm for Upper Limb Prosthesis Control: Validation in Real-Life Scenarios on Amputees	(10.1109/TNSRE.2014.2361478)	\\																			
S61	&	Shamsollahi, P.; Cheung, K. W.; Quan Chen  \&  Germain, E. H.	&	2001	&	A neural network based very short term load forecaster for the interim ISO New England electricity market system	(10.1109/PICA.2001.932351)	\\																			
S62	&	Mariani, L.; Pezze, M.; Riganelli, O.  \&  Santoro, M.	&	2012	&	AutoBlackTest: Automatic Black-Box Testing of Interactive Applications	(10.1109/ICST.2012.88)	\\																			
S63	&	Renaudie, D.; Zuluaga, M. A.  \&  Acuna-Agost, R.	&	2018	&	Benchmarking Anomaly Detection Algorithms in an Industrial Context: Dealing with Scarce Labels and Multiple Positive Types	(10.1109/BigData.2018.8621956)	\\																			
S64	&	Sayed, B.; Traoré, I.; Woungang, I.  \&  Obaidat, M. S.	&	2013	&	Biometric Authentication Using Mouse Gesture Dynamics	(10.1109/JSYST.2012.2221932)	\\																			
S65	&	Xi-Zheng Zhang	&	2007	&	Building Personalized Recommendation System in E-Commerce using Association Rule-Based Mining and Classification	(10.1109/ICMLC.2007.4370866)	\\																			
S66	&	Vamsikrishna, K. M.; Dogra, D. P.  \&  Desarkar, M. S.	&	2016	&	Computer-Vision-Assisted Palm Rehabilitation With Supervised Learning	(10.1109/TBME.2015.2480881)	\\																			
S67	&	Wang, W.; Zeng, Z.; Ding, W.; Yu, H.  \&  Rose, H.	&	2019	&	Concept and Validation of a Large-scale Human-machine Safety System Based on Real-time UWB Indoor Localization*	(10.1109/IROS40897.2019.8968572)	\\																			
S68	&	Teleka, R.; Green, J.  \&  Dickens, J.	&	2013	&	CSIR center for mining innovation and the simulated test stope	(10.1109/AFRCON.2013.6757795)	\\																			
S69	&	He, M.  \&  He, D.	&	2017	&	Deep Learning Based Approach for Bearing Fault Diagnosis	(10.1109/TIA.2017.2661250)	\\																			
S70	&	Wang, H.; Ruan, J.; Wang, G.; Zhou, B.; Liu, Y.; Fu, X.  \&  Peng, J.	&	2018	&	Deep Learning-Based Interval State Estimation of AC Smart Grids Against Sparse Cyber Attacks	(10.1109/TII.2018.2804669)	\\																			
S71	&	Aykut, T.; Karimi, M.; Burgmair, C.; Finkenzeller, A.; Bachhuber, C.  \&  Steinbach, E.	&	2018	&	Delay Compensation for a Telepresence System With 3D 360 Degree Vision Based on Deep Head Motion Prediction and Dynamic FoV Adaptation	(10.1109/LRA.2018.2864359)	\\																			
S72	&	Grammatikopoulou, M.; Zhang, L.  \&  Yang, G.	&	2018	&	Depth Estimation of Optically Transparent Microrobots Using Convolutional and Recurrent Neural Networks	(10.1109/IROS.2018.8593776)	\\																			
S73	&	Gill, V. S.; Harris, A. C.; Swami, S. G.  \&  Conrad, J. M.	&	2012	&	Design, development and validation of sensors for a Simulation Environment for Autonomous Robots	(10.1109/SECon.2012.6196926)	\\																			
S74	&	Shadrin, D.; Menshchikov, A.; Ermilov, D.  \&  Somov, A.	&	2019	&	Designing Future Precision Agriculture: Detection of Seeds Germination Using Artificial Intelligence on a Low-Power Embedded System	(10.1109/JSEN.2019.2935812)	\\																			
S75	&	Zhang, Y.; Guo, L.; Gao, B.; Qu, T.  \&  Chen, H.	&	2020	&	Deterministic Promotion Reinforcement Learning Applied to Longitudinal Velocity Control for Automated Vehicles	(10.1109/TVT.2019.2955959)	\\																			
S76	&	Suzuki, S.; Amemiya, Y.  \&  Sato, M.	&	2019	&	Enhancement of gross-motor action recognition for children by CNN with OpenPose	(10.1109/IECON.2019.8927828)	\\																			
S77	&	Ryota Kurozumi; Kosuke Tsuji; Shin-ichi Ito; Katsuya Sato; Shoichiro Fujisawa  \&  Toru Yamamoto	&	2010	&	Experimental validation of an online adaptive and learning obstacle avoiding support system for the electric wheelchairs	(10.1109/ICSMC.2010.5642211)	\\																			
S78	&	Li, N.; Oyler, D. W.; Zhang, M.; Yildiz, Y.; Kolmanovsky, I.  \&  Girard, A. R.	&	2018	&	Game Theoretic Modeling of Driver and Vehicle Interactions for Verification and Validation of Autonomous Vehicle Control Systems	(10.1109/TCST.2017.2723574)	\\																			
S79	&	Liu, K.; Li, Y.; Hu, X.; Lucu, M.  \&  Widanage, W. D.	&	2020	&	Gaussian Process Regression With Automatic Relevance Determination Kernel for Calendar Aging Prediction of Lithium-Ion Batteries	(10.1109/TII.2019.2941747)	\\																			
S80	&	Heydarzadeh, M.; Kia, S. H.; Nourani, M.; Henao, H.  \&  Capolino, G.	&	2016	&	Gear fault diagnosis using discrete wavelet transform and deep neural networks	(10.1109/IECON.2016.7793549)	\\																			
S81	&	Livanos, G.; Zervakis, M.; Chalkiadakis, V.; Moirogiorgou, K.; Giakos, G.  \&  Papandroulakis, N.	&	2018	&	Intelligent Navigation and Control of a Prototype Autonomous Underwater Vehicle for Automated Inspection of Aquaculture net pen cages	(10.1109/IST.2018.8577180)	\\																			
S82	&	Zhou, Y.; Yang, J.  \&  Zheng, L.	&	2019	&	Multi-Agent Based Hyper-Heuristics for Multi-Objective Flexible Job Shop Scheduling: A Case Study in an Aero-Engine Blade Manufacturing Plant	(10.1109/ACCESS.2019.2897603)	\\																			
S83	&	Kang, C. M.; Lee, S.  \&  Chung, C. C.	&	2018	&	Multirate Lane-Keeping System With Kinematic Vehicle Model	(10.1109/TVT.2018.2864329)	\\																			
S84	&	Yang, L.; Tuzel, O.; Chen, W.; Meer, P.; Salaru, G.; Goodell, L. A.  \&  Foran, D. J.	&	2009	&	PathMiner: A Web-Based Tool for Computer-Assisted Diagnostics in Pathology	(10.1109/TITB.2008.2008801)	\\																			
S85	&	Jung, S.; Hwang, S.; Shin, H.  \&  Shim, D. H.	&	2018	&	Perception, Guidance, and Navigation for Indoor Autonomous Drone Racing Using Deep Learning	(10.1109/LRA.2018.2808368)	\\																			
S86	&	Jo, K.; Jo, Y.; Suhr, J. K.; Jung, H. G.  \&  Sunwoo, M.	&	2015	&	Precise Localization of an Autonomous Car Based on Probabilistic Noise Models of Road Surface Marker Features Using Multiple Cameras	(10.1109/TITS.2015.2450738)	\\																			
S87	&	Jonsson, L.; Broman, D.; Sandahl, K.  \&  Eldh, S.	&	2012	&	Towards Automated Anomaly Report Assignment in Large Complex Systems Using Stacked Generalization	(10.1109/ICST.2012.124)	\\																			
S88	&	Souza, K. E. S.; Seruffo, M. C. R.; De Mello, H. D.; Souza, D. D. S.  \&  Vellasco, M. M. B. R.	&	2019	&	User Experience Evaluation Using Mouse Tracking and Artificial Intelligence	(10.1109/ACCESS.2019.2927860)	\\																			
S89	&	Fei, X.; Zhang, Q.  \&  Ling, Q.	&	2019	&	Vehicle Exhaust Concentration Estimation Based on an Improved Stacking Model	(10.1109/ACCESS.2019.2958703)	\\																			
S90	&	Khosrowjerdi, H.; Meinke, K.  \&  Rasmusson, A.	&	2018	&	Virtualized-Fault Injection Testing: A Machine Learning Approach	(10.1109/ICST.2018.00037)	\\																			
\hline																											
\end{supertabular}																											
\label{tab:PrimaryStudies}

\onecolumn																											
\tablefirsthead{ID & Domain (F4) & Task (F5) & Complexity (F6) & Malfunction (F7) & Validation (F8) & Continuous validation (F9) & ML (F12) \\ \hline}																											
\tablehead{\multicolumn{4}{l}{\footnotesize\sl Table \ref{tab:FullData} continued}\\ ID & Domain (F4) & Task (F5) & Complexity (F6) & Malfunction (F7) & Validation (F8) & Continuous validation (F9)  & ML (F12) \\ \hline}																											
\tablecaption{The full data to the research questions from the primary studies. The columns are the fields of the data extraction form (Section \ref{sec:InformationExtraction}).}																											
\footnotesize																											
\begin{supertabular}{p{0.2cm}p{2.1cm}p{2cm}p{2cm}p{2cm}p{2cm}p{3.1cm}p{2cm}}

S1	&	car	&	transportation	&	multi-component	&	lethal	&	simulation	&	no - test description	&	plausible	\\												
S2	&	robot	&	maintenance	&	multi-component	&	nuisance	&	trial	&	no	&	yes	\\												
S3	&	wearable ai	&	recognition	&	system	&	bias	&	trial	&	no	&	yes	\\												
S4	&	smart environment	&	security	&	system	&	privacy	&	trial	&	no	&	yes	\\												
S5	&	car	&	transportation	&	multi-component	&	lethal	&	simulation	&	failure monitor	&	yes	\\												
S6	&	car	&	transportation	&	multi-component	&	lethal	&	simulation	&	safety channel	&	yes	\\												
S7	&	medical	&	recognition	&	system	&	bias	&	trial	&	no	&	yes	\\												
S8	&	robot	&	care	&	multi-component	&	nuisance	&	simulation \& trial	&	no	&	yes	\\												
S9	&	robot	&	unspecified	&	system	&	mission critical	&	simulation	&	no - test description	&	plausible	\\												
S10	&	industrial	&	recognition	&	model	&	mission critical	&	simulation	&	no	&	yes	\\												
S11	&	industrial	&	recognition	&	system	&	economic	&	trial	&	no	&	yes	\\												
S12	&	commercial	&	recognition	&	model	&	economic	&	statistical proof	&	no	&	yes	\\												
S13	&	robot	&	recognition	&	system	&	nuisance	&	trial	&	no	&	yes	\\												
S14	&	testing	&	recognition	&	system	&	bias	&	trial	&	no	&	yes	\\												
S15	&	car	&	transportation	&	multi-component	&	lethal	&	trial	&	redundancy	&	yes	\\												
S16	&	robot	&	recognition	&	system	&	nuisance	&	trial	&	no	&	yes	\\												
S17	&	car	&	transportation	&	multi-component	&	lethal	&	simulation	&	no	&	yes	\\												
S18	&	commercial	&	recognition	&	model	&	economic	&	trial	&	no	&	yes	\\												
S19	&	robot	&	maintenance	&	multi-component	&	mission critical	&	simulation	&	no	&	yes	\\												
S20	&	robot	&	assembly	&	multi-component	&	economic	&	simulation \& trial	&	voting	&	yes	\\												
S21	&	commercial	&	recognition	&	system	&	nuisance	&	trial	&	no	&	yes	\\												
S22	&	robot	&	critical missions	&	multi-component	&	lethal	&	simulation \& trial	&	no - test description	&	yes	\\												
S23	&	unspecified	&	unspecified	&	unspecified	&	unspecified	&	trial	&	no - test description	&	yes	\\												
S24	&	agriculture	&	recognition	&	system	&	economic	&	trial	&	no	&	yes	\\												
S25	&	wearable ai	&	decision support	&	system	&	nuisance	&	trial	&	no	&	yes	\\												
S26	&	robot	&	search and rescue	&	multi-component	&	lethal	&	trial	&	failure monitor	&	no	\\												
S27	&	robot	&	maintenance	&	system	&	economic	&	trial	&	no	&	yes	\\												
S28	&	robot	&	maintenance	&	multi-component	&	economic	&	simulation	&	no - test description	&	yes	\\												
S29	&	robot	&	care	&	multi-component	&	nuisance	&	trial	&	failure monitor	&	yes	\\												
S30	&	commercial	&	recognition	&	model	&	economic	&	trial	&	no	&	yes	\\												
S31	&	industrial	&	recognition	&	model	&	economic	&	trial	&	no	&	yes	\\												
S32	&	industrial	&	recognition	&	system	&	environmental	&	trial	&	no	&	yes	\\												
S33	&	commercial	&	recognition	&	model	&	economic	&	trial	&	no	&	yes	\\												
S34	&	robot	&	loading	&	multi-component	&	nuisance	&	simulation	&	no - test description	&	no	\\												
S35	&	robot	&	unspecified	&	multi-component	&	mission critical	&	simulation \& trial	&	safety channel	&	yes	\\												
S36	&	commercial	&	decision support	&	model	&	unspecified	&	trial	&	no	&	yes	\\												
S37	&	smart environment	&	safety	&	multi-component	&	lethal	&	simulation	&	no - test description	&	no	\\												
S38	&	industrial	&	recognition	&	system	&	economic	&	simulation \& trial	&	no	&	plausible	\\												
S39	&	unspecified	&	decision support	&	system	&	economic	&	expert opinion	&	no - test description	&	yes	\\												
S40	&	robot	&	unspecified	&	multi-component	&	nuisance	&	trial	&	no	&	yes	\\												
S41	&	wearable ai	&	recognition	&	system	&	lethal	&	expert opinion	&	no	&	yes	\\												
S42	&	aviation	&	recognition	&	multi-component	&	nuisance	&	simulation	&	no	&	yes	\\												
S43	&	car	&	recognition	&	model	&	economic	&	trial	&	no	&	yes	\\												
S44	&	smart environment	&	safety	&	system	&	lethal	&	simulation	&	no - test description	&	plausible	\\												
S45	&	robot	&	loading	&	multi-component	&	mission critical	&	simulation	&	failure monitor	&	yes	\\												
S46	&	medical	&	unspecified	&	model	&	unspecified	&	statistical proof	&	no - test description	&	no	\\												
S47	&	industrial	&	design	&	model	&	economic	&	trial	&	no	&	yes	\\												
S48	&	robot	&	loading	&	multi-component	&	harmful	&	simulation \& trial	&	safety channel	&	yes	\\												
S49	&	industrial	&	recognition	&	model	&	economic	&	statistical proof	&	no	&	yes	\\												
S50	&	robot	&	transportation	&	multi-component	&	lethal	&	simulation \& trial	&	output restrictions	&	yes	\\												
S51	&	medical	&	assesment	&	system	&	harmful	&	statistical proof	&	no	&	yes	\\												
S52	&	agriculture	&	recognition	&	system	&	economic	&	trial	&	no	&	yes	\\												
S53	&	government	&	control	&	model	&	harmful	&	simulation	&	no	&	yes	\\												
S54	&	gaming	&	testing	&	multi-component	&	economic	&	trial	&	no	&	yes	\\												
S55	&	commercial	&	decision support	&	model	&	economic	&	statistical proof	&	no	&	yes	\\												
S56	&	unspecified	&	unspecified	&	unspecified	&	unspecified	&	simulation	&	no - test description	&	yes	\\												
S57	&	commercial	&	recognition	&	model	&	economic	&	statistical proof	&	no	&	yes	\\												
S58	&	testing	&	recognition	&	system	&	economic	&	trial	&	no	&	yes	\\												
S59	&	wearable ai	&	recognition	&	system	&	nuisance	&	trial	&	no	&	yes	\\												
S60	&	medical	&	recognition	&	multi-component	&	nuisance	&	trial	&	no	&	yes	\\												
S61	&	industrial	&	decision support	&	model	&	mission critical	&	statistical proof	&	no	&	yes	\\												
S62	&	testing	&	testing	&	system	&	economic	&	trial	&	no	&	yes	\\												
S63	&	industrial	&	recognition	&	model	&	unspecified	&	trial	&	no	&	yes	\\												
S64	&	safety	&	recognition	&	system	&	privacy	&	trial	&	no	&	yes	\\												
S65	&	commercial	&	decision support	&	system	&	nuisance	&	trial	&	no	&	yes	\\												
S66	&	commercial	&	rehabilitation	&	multi-component	&	harmful	&	trial	&	no	&	yes	\\												
S67	&	robot	&	security	&	multi-component	&	harmful	&	trial	&	safety channel	&	no	\\												
S68	&	safety	&	safety	&	multi-component	&	lethal	&	simulation	&	no - test description	&	plausible	\\												
S69	&	industrial	&	maintenance	&	system	&	economic	&	statistical proof	&	no	&	yes	\\												
S70	&	commercial	&	safety	&	model	&	harmful	&	simulation	&	no	&	yes	\\												
S71	&	wearable ai	&	control	&	system	&	nuisance	&	trial	&	no	&	yes	\\												
S72	&	robot	&	recognition	&	system	&	mission critical	&	simulation	&	no	&	yes	\\												
S73	&	robot	&	control	&	system	&	unspecified	&	simulation	&	no - test description	&	plausible	\\												
S74	&	agriculture	&	recognition	&	system	&	economic	&	trial	&	input restrictions	&	yes	\\												
S75	&	car	&	control	&	system	&	lethal	&	simulation \& trial	&	no	&	yes	\\												
S76	&	medical	&	assesment	&	system	&	harmful	&	statistical proof	&	no	&	yes	\\												
S77	&	medical	&	assistance	&	system	&	harmful	&	simulation \& trial	&	no	&	yes	\\												
S78	&	car	&	transportation	&	multi-component	&	lethal	&	simulation	&	safety channel \& output restrictions	&	yes	\\												
S79	&	commercial	&	assesment	&	model	&	nuisance	&	statistical proof	&	no	&	yes	\\												
S80	&	safety	&	assesment	&	system	&	economic	&	trial	&	no	&	yes	\\												
S81	&	agriculture	&	control	&	multi-component	&	economic	&	trial	&	no	&	plausible	\\												
S82	&	industrial	&	scheduling	&	model	&	economic	&	statistical proof \& expert opinion \& trial	&	no	&	yes	\\												
S83	&	car	&	control	&	multi-component	&	lethal	&	simulation \& trial	&	no	&	yes	\\												
S84	&	medical	&	recognition	&	system	&	lethal	&	trial	&	no	&	yes	\\												
S85	&	aviation	&	transportation	&	multi-component	&	mission critical	&	statistical proof \& trial	&	no	&	yes	\\												
S86	&	car	&	transportation	&	multi-component	&	lethal	&	simulation \& trial	&	no	&	no	\\												
S87	&	industrial	&	recognition	&	model	&	economic	&	statistical proof	&	input restrictions	&	yes	\\												
S88	&	commercial	&	assesment	&	system	&	nuisance	&	trial	&	no	&	yes	\\												
S89	&	smart environment	&	assesment	&	model	&	nuisance	&	statistical proof	&	no	&	yes	\\												
S90	&	testing	&	testing	&	system	&	economic	&	trial	&	no	&	yes	\\												
\hline																											
\end{supertabular}																											
\label{tab:FullData}

\twocolumn

\tablefirsthead{	&	Context	&	Design	&	Validity	&	Rigor	&	Relevance	\\ \hline}																								
\tablehead{\multicolumn{4}{l}{\footnotesize\sl Table \ref{tab:QAData} continued}\\	&	Context	&	Design	&	Validity	&	Rigor	&	Relevance	\\ \hline}																								
\tablecaption{ Quality assessment and evidence levels as a measures for rigor and relevance.}																																			
\footnotesize																																			
\begin{supertabular}{p{1cm}>{\centering}p{1cm}>{\centering}p{1cm}>{\centering}p{1cm}>{\centering}p{1cm} c}																																			
S1	&	1	&	1	&	0	&	2	&	4	 \\																								
S2	&	0.5	&	1	&	0	&	1.5	&	4	 \\																								
S3	&	0.5	&	1	&	0	&	1.5	&	4	 \\																								
S4	&	1	&	1	&	0	&	2	&	4	 \\																								
S5	&	1	&	1	&	0	&	2	&	4	 \\																								
S6	&	1	&	0.5	&	0	&	1.5	&	4	 \\																								
S7	&	1	&	1	&	1	&	3	&	4	 \\																								
S8	&	1	&	1	&	0.5	&	2.5	&	5	 \\																								
S9	&	0	&	1	&	0	&	1	&	4	 \\																								
S10	&	1	&	0.5	&	0	&	1.5	&	4	 \\																								
S11	&	0.5	&	0.5	&	0	&	1	&	4	 \\																								
S12	&	0	&	1	&	0.5	&	1.5	&	4	 \\																								
S13	&	1	&	0.5	&	0	&	1.5	&	4	 \\																								
S14	&	0.5	&	0.5	&	0.5	&	1.5	&	4	 \\																								
S15	&	1	&	1	&	0	&	2	&	4	 \\																								
S16	&	0.5	&	0	&	0	&	0.5	&	4	 \\																								
S17	&	0.5	&	0.5	&	0.5	&	1.5	&	4	 \\																								
S18	&	0.5	&	1	&	0	&	1.5	&	4	 \\																								
S19	&	1	&	0.5	&	0	&	1.5	&	4	 \\																								
S20	&	1	&	1	&	0	&	2	&	4	 \\																								
S21	&	1	&	0.5	&	0	&	1.5	&	4	 \\																								
S22	&	1	&	1	&	0	&	2	&	4	 \\																								
S23	&	0.5	&	1	&	0	&	1.5	&	4	 \\																								
S24	&	1	&	1	&	0	&	2	&	4	 \\																								
S25	&	1	&	1	&	1	&	3	&	4	 \\																								
S26	&	1	&	1	&	0	&	2	&	6	 \\																								
S27	&	0.5	&	0.5	&	0	&	1	&	4	 \\																								
S28	&	0.5	&	0.5	&	0	&	1	&	4	 \\																								
S29	&	0.5	&	0.5	&	0	&	1	&	5	 \\																								
S30	&	1	&	1	&	1	&	3	&	6	 \\																								
S31	&	1	&	0.5	&	0	&	1.5	&	5	 \\																								
S32	&	1	&	1	&	0	&	2	&	4	 \\																								
S33	&	1	&	0	&	0	&	1	&	5	 \\																								
S34	&	1	&	1	&	0	&	2	&	4	 \\																								
S35	&	1	&	1	&	0	&	2	&	4	 \\																								
S36	&	1	&	0	&	0	&	1	&	5	 \\																								
S37	&	1	&	0.5	&	0	&	1.5	&	4	 \\																								
S38	&	1	&	1	&	0	&	2	&	5	 \\																								
S39	&	1	&	1	&	0	&	2	&	4	 \\																								
S40	&	1	&	1	&	0	&	2	&	4	 \\																								
S41	&	1	&	1	&	0	&	2	&	4	 \\																								
S42	&	1	&	1	&	0	&	2	&	4	 \\																								
S43	&	1	&	0.5	&	0	&	1.5	&	4	 \\																								
S44	&	1	&	0	&	0	&	1	&	4	 \\																								
S45	&	0.5	&	0.5	&	0.5	&	1.5	&	4	 \\																								
S46	&	1	&	1	&	0	&	2	&	4	 \\																								
S47	&	1	&	1	&	0	&	2	&	5	 \\																								
S48	&	1	&	0.5	&	0	&	1.5	&	5	 \\																								
S49	&	1	&	0.5	&	0	&	1.5	&	4	 \\																								
S50	&	1	&	1	&	0	&	2	&	4	 \\																								
S51	&	1	&	0.5	&	0	&	1.5	&	5	 \\																								
S52	&	1	&	1	&	0	&	2	&	4	 \\																								
S53	&	1	&	0.5	&	0	&	1.5	&	4	 \\																								
S54	&	1	&	1	&	1	&	3	&	5	 \\																								
S55	&	0.5	&	0.5	&	0	&	1	&	4	 \\																								
S56	&	1	&	0.5	&	0	&	1.5	&	5	 \\																								
S57	&	1	&	1	&	0	&	2	&	5	 \\																								
S58	&	1	&	0.5	&	0	&	1.5	&	5	 \\																								
S59	&	0.5	&	1	&	0	&	1.5	&	4	 \\																								
S60	&	1	&	1	&	0	&	2	&	4	 \\																								
S61	&	0.5	&	1	&	0	&	1.5	&	5	 \\																								
S62	&	0.5	&	0.5	&	1	&	2	&	5	 \\																								
S63	&	0.5	&	0.5	&	1	&	2	&	4	 \\																								
S64	&	1	&	1	&	0	&	2	&	4	 \\																								
S65	&	0.5	&	1	&	0	&	1.5	&	5	 \\																								
S66	&	1	&	1	&	0	&	2	&	4	 \\																								
S67	&	0.5	&	0.5	&	0	&	1	&	5	 \\																								
S68	&	1	&	1	&	0	&	2	&	5	 \\																								
S69	&	1	&	1	&	0	&	2	&	4	 \\																								
S70	&	1	&	0.5	&	0	&	1.5	&	4	 \\																								
S71	&	1	&	1	&	0.5	&	2.5	&	4	 \\																								
S72	&	1	&	1	&	0	&	2	&	4	 \\																								
S73	&	0.5	&	1	&	0	&	1.5	&	4	 \\																								
S74	&	1	&	0.5	&	0	&	1.5	&	4	 \\																								
S75	&	1	&	1	&	0	&	2	&	4	 \\																								
S76	&	1	&	1	&	0	&	2	&	4	 \\																								
S77	&	1	&	1	&	0	&	2	&	4	 \\																								
S78	&	1	&	1	&	0.5	&	2.5	&	4	 \\																								
S79	&	0.5	&	1	&	0	&	1.5	&	4	 \\																								
S80	&	1	&	1	&	0	&	2	&	4	 \\																								
S81	&	1	&	0	&	0	&	1	&	4	 \\																								
S82	&	1	&	0.5	&	0	&	1.5	&	6	 \\																								
S83	&	1	&	0.5	&	0	&	1.5	&	4	 \\																								
S84	&	1	&	1	&	0	&	2	&	5	 \\																								
S85	&	1	&	1	&	0	&	2	&	4	 \\																								
S86	&	1	&	1	&	0	&	2	&	4	 \\																								
S87	&	0.5	&	1	&	0	&	1.5	&	5	 \\																								
S88	&	0.5	&	1	&	0	&	1.5	&	5	 \\																								
S89	&	0.5	&	1	&	0	&	1.5	&	5	 \\																								
S90	&	0.5	&	1	&	0	&	1.5	&	4	 \\																								
																																			
\hline																																			
\end{supertabular}																																			
\label{tab:QAData}																																			

\twocolumn

\bibliographystyle{elsarticle-num}
\bibliography{slrref}

\end{document}